\documentclass[twocolumn,prb,aps,epsfig,amsmath,showpacs]{revtex4}
\usepackage{epsfig}
\usepackage{dcolumn}
\usepackage{bm}
\usepackage{color}
\begin{document}
\title{Metal-insulator transitions in GdTiO$_3$/SrTiO$_3$ superlattices}
\author{Se Young Park}
\thanks{Current address: Department of Physics and Astronomy, Rutgers University, Piscataway, NJ 08854, USA}
\author{Andrew J. Millis}
\affiliation{Department of Physics, Columbia University, New York, New York 10027, USA}


\begin{abstract}
The density functional plus U method is used to obtain the electronic structure, lattice relaxation and metal-insulator phase diagram of superlattices consisting of $m$ layers of Gadolinium Titanate (GdTiO$_{3}$) alternating with $n$ layers of Strontium Titanate (SrTiO$_{3}$). Metallic phases occur when the number of SrTiO$_3$ layers is large or the interaction $U$ is small. In metallic phases, the mobile electrons are found in the SrTiO$_3$ layers, with  near-interface electrons occupying $xy$-derived bands, while away from the interface the majority of electrons reside in $xz/yz$  bands. As the thickness of the SrTiO$_3$ layers decreases or the on-site interaction U increases a metal-insulator transition occurs.  Two different insulating states are found. When the number of SrTiO$_{3}$ layers is larger than one, we find an insulating state with two sublattice charge and orbital ordering and associated Ti-O bond length disproportionations. When the number of SrTiO$_{3}$ units per layer is one, a different insulating phase occurs driven by orbital ordering within the quasi one-dimensional $xz/yz$ bonding bands connecting Ti atoms across the SrO layer. In this phase there is no sublattice charge ordering or bond disproportionation. The critical U for the single-layer  insulator is $\sim$ 2.5 eV, much less than  critical U $\sim$ 3.5 eV required to drive the metal-insulator transition when the number of SrTiO$_3$ is larger than one.  Inconsistencies between the calculation and the experiment suggest that many-body correlations may be important. A local inversion symmetry breaking around Ti atoms suggests the possibility of in-plane ferroelectric polarization in the insulating phase. 
\end{abstract}

\pacs{73.20.-r, 71.30.+h, 75.70.-i, 71.27.+a} 
\maketitle

\section{Introduction}
Complex oxide systems have been investigated extensively because of the variety of interesting phases that may be achieved by alloying and pressure\cite{Imada98}. Of particular interest are materials with structures based on the ABO$_3$ perovskite motif (here A represents a rare earth or alkali atom, B a transition metal, and O an oxygen), because the many available possibilities for the A and B site ion lead to a great diversity of interesting properties. Over the last decade it has become clear that complex oxide heterostructures may be grown with precise layer-by-layer control of the composition \cite{Ohtomo02,Ohtomo04,Moetakef11,Moetakef11-2} enabling the creation of what are in effect new materials with potentially new properties. Moreover, carrier doping by charge transfer across interfaces can be much higher than that obtainable by alloying or vacancy formation in bulk materials\cite{Moetakef11,Biscaras10,Seo07} while electronic gating allows the continuous control of carrier doping.\cite{Ahn03,Hwang12,Zubko11,Caviglia08,Mannhart08}. Heterostructures comprised of two different ABO$_{3}$  transition metal oxide perovskites with comparable bulk lattice constants have been reported to exhibit metal-insulator transitions (MIT)\cite{Thiel06,Yoshimatsu10,Perucchi10,Son10},  magnetism\cite{Luders09,Ariando11,Brinkman07,Bert11} as well as superconductivity.\cite{Biscaras10,Reyren07}

The electronic properties of semiconductor heterostructures are determined by bulk band gaps, work functions, and the position of donor (acceptor) levels. In many transition metal oxides, electronic properties are, in addition, sensitive  to  structural features including octahedral rotation angles and transition-metal-oxygen bond lengths. Thus local structural changes across interfaces may play an important role in determining the electronic properties of oxide heterostructures. Moreover structural distortions propagate only a few lattice constants, so superlattice thickness has a structural as well as a quantum confinement effect. Undertanding the interplay of the various factors that contribute to metal-insulator transitions in oxide superlattices is an important open problem.

Heterostructures comprised of Mott-insulating GdTiO$_{3}$(GTO) and band insulating SrTiO$_{3}$(STO) provide an interesting model system. A single interface separating semi-infinite slabs of GdTiO$_{3}$ and SrTiO$_{3}$ is found to be metallic with sheet charge density about a half electron per in-plane unit cell,\cite{Moetakef11} consistent with elementary `polar catastrophe' notions of interface doping and in contrast with the more complicated behavior of the widely studied LaAlO$_3$(LAO) systems, where the density of mobile carriers is much less than the polar catastrophe value \cite{Thiel06}.  Part of the issue may be that GdTiO$_3$ and SrTiO$_3$ share a common Ti$-$O network so that the main defects at an interface would be Gd$-$Sr antisite defects, which might be expected to disrupt the electron structure less than the Al$-$Ti antisite defects that might occur at the LAO-STO interface. A further interesting feature  is that GdTiO$_3$ is itself a Mott insulator, and has a large amplitude rotational distortion away from the basic cubic ABO$_3$ perovskite structure. Taken together, these features suggest  that the GTO$-$STO system may exhibit an interesting interplay of structural, interface, quantum confinement, and correlation effects, perhaps not too badly complicated by disorder, making it a suitable model system for investigation of general issues relating to metal-insulator transitions in oxide heterostructures. 

Recently, thickness-dependent metal insulator transitions have been reported in GTO$-$STO heterostructures.\cite{Moetakef11-2}  As the thickness of SrTiO$_{3}$ decreases to two unit cells of SrTiO$_{3}$, the interface is found to become insulating. Both in metallic and in insulating interfaces, variations of the crystal structure that decay within several unit-cells of the interface are observed.\cite{Zhang13} In metallic interfaces, quantum oscillation measurements indicate that the conducting carriers move primarily in the plane of the interface.\cite{Moetakef12-2} The theory of the insulating phase was investigated on the basis of first-principles DFT+U calculations by Chen and Balents,\cite{Chen13} who proposed a novel dimerization mechanism that could lead to  an insulating ground state when the number of SrTiO$_{3}$ layers, $n=1$ and by Bjaalie {\it et al.}\cite{Bjaalie14} who reported  an insulating ground state with $n=2$ using DFT with a hybrid functional. In this paper we present a more general DFT+U investigation of the metal-insulator phase diagram and structural properties of the GTO-STO system.  Although DFT+U is a Hartree approximation which  does not capture the full complexity of many-body physics, it enables the investigation of the interplay between structural relaxation and electron correlation effects.  The changes in the electronic structure in different structural phases are investigated and the spatial distribution of the electron gas and its orbital polarization for different thickness of SrTiO$_{3}$ are determined as function of interaction strength, $U$. 

 \begin{figure}[htbp]
\begin{center}
\includegraphics[width=1\columnwidth, angle=-0]{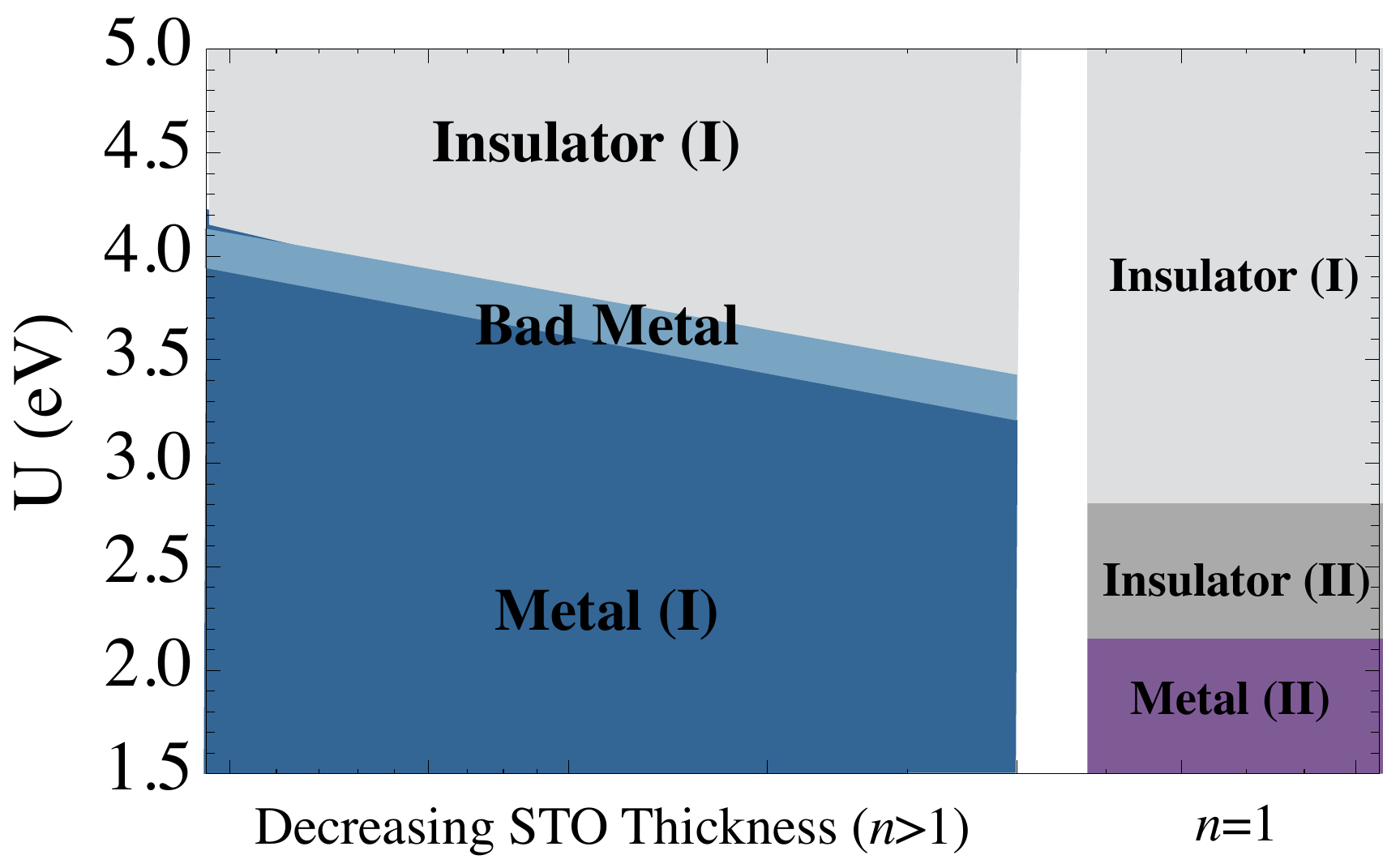}
\caption{(color online) Schematic phase diagram of electronic phases as a function of SrTiO$_{3}$ layer thickness $n$ and on-site Coulomb repulsion U. }
\label{fig:electphase}
\end{center}
\end{figure}

Our crucial finding, illustrated in Fig.~\ref{fig:electphase},  is that there are two different insulating phases, one occurring for ultra-thin ($n=1$) and  one for thicker ($n>1$) STO layers. In the DFT+U approximation employed here, some form of charge, orbital, or magnetic ordering is required to stabilize insulating phases.  When there is more than one SrTiO$_{3}$ unit cell ($n>1$), we find a MIT from ferromagnetic metal to charge-ordered insulator with a narrow intermediate regime of charge ordered low density of states behavior, which we interpret as the signature in this calculations of a ``bad metal'' phase.\cite{Gunnarsson03,Hussey04} The metal-insulator transition is accompanied by a structural transition resulting in sublattice bond disproportionation, and within DFT+U the  charge ordering is the main driving force for insulating/bad-metallic behavior. In the case of a single SrO layer ($n=1$) we find two insulating phases. In one of them,  $yz/xz$ orbitals on opposite sides of the SrO layer form bonding and antibonding states; the antibonding states are empty and the spin-polarized half-filled bonding bands become gapped by orbital ordering. This transition from ferromagnetic metal to ferromagnetic insulator phase is consistent with the  previous Hartree-Fock calculations of Chen and Balents.\cite{Chen13} The critical interaction strength required to drive this insulating phase is significantly lower than that needed to drive the insulating phase I. As on-site Coulomb interaction increases at  $n=1$ a transition occurs to a different insulating phase, characterized by two-sublattice charge order and similar to the insulating phase found at $n>1$.

The rest of the paper is organized as follows. In section \ref{sec:method} we state the calculation methods that we use. Section \ref{sec:str} presents structural phase transitions in the different crystal structures. Electronic structures in each phase are are presented in Section \ref{sec:twolayer}. Section \ref{sec:onelayer} is devoted to an analysis of the electronic structure with a single SrO layer. In Section \ref{sec:PT} we present structural and electronic phase diagram in terms of layer thickness and electron correlation. Section \ref{sec:invsym} presents  possible ferroelectric polarization from  broken inversion symmetry and centrosymmetry. In Section \ref{sec:ba} we discuss a simplified view of metal insulator transition based on band alignment, followed by summary in Sec, \ref{sec:sum}.

\section{Calculational Methods}
\label{sec:method}
The electronic structure and atomic structures are calculated using the GGA+U method as implemented in the Vienna {\it ab-initio} simulation package (VASP).\cite{Kresse96,Kresse99} We use a plane wave basis set with energy cutoff 500 eV and the projected augmented wave method. For all superlattices we use a unit cell  $\sqrt{2}\times\sqrt{2}\times 2l$ ($l:$ integer) structure to describe proper octahedral rotations and orbital ordering of the superlattices. All the structures are fully relaxed while the in-plane lattice constant is constrained to the substrate ($a=b=3.86\AA$). Convergence is reached if the energy difference between two consecutive iterations is within 0.1 meV for electronic iterations and 1 meV for ionic relaxations. The minimum $k$-point grid size is $6\times 6 \times 2$. For on-site Coulomb interaction the rotationally invariant form\cite{Liechtenstein95} is used for Ti-$d$ orbitals. We treat the Hubbard $U$ as an adjustable parameter while fixing the value of $J$ as 0.6 eV.

\section{Lattice Structures}
\label{sec:str}

\begin{figure}[htbp]
\begin{center}
\includegraphics[width=0.8\columnwidth, angle=-0]{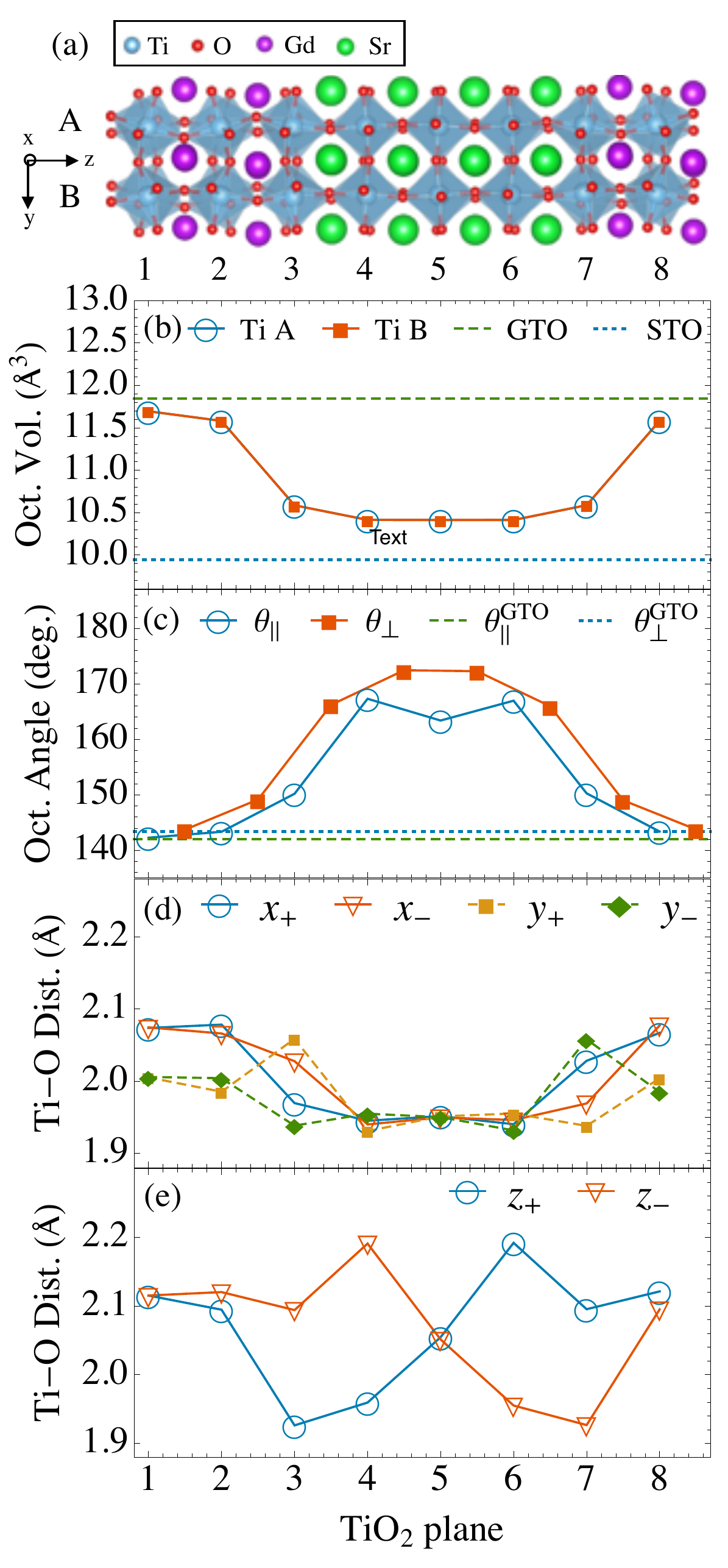}
\caption{(color online) Structural properties of (GTO)$_{4}$(STO)$_{4}$ superlattices in the metallic phase of the superlattice computed as described in the main text with $U=3$ eV. (a) The atomic structure of the superlattice. A and B represent two inequivalent in-plane Ti atoms and the numbers in the bottom label the TiO$_{2}$ planes. (b) Octahedral volumes around A and B sublattice Ti atoms. The green dashed and blue dotted lines are respectively the octahedral volumes of bulk GTO and STO calculated with experimental in-plane lattice constant. (c) Average values of in-plane ($\theta_{||}$) and out-of-plane ($\theta_{\perp}$) Ti-O-Ti angles.  The green dashed and blue dotted lines are the in-plane and out-of-plane octahedral angle calculated for bulk GTO with experimental substrate, respectively. (d) In-plane (x, y) Ti-O bond lengths of A sublattice Ti atoms. (e) Out of plane (z) bond lengths of A sublattice Ti atoms.  The six oxygen atoms around the each Ti atom are labeled by the direction from Ti to oxygen atoms expressed by $x_{\pm}, y_{\pm},$ and $z_{\pm}$ where the direction ($x,y,z$) and subscripted sign represent pseudo-cubic axis defined in panel (a) and plus and minus direction, respectively. Exchanging $x_{\pm}$ to $y_{\mp}$ gives the B sublattice Ti-O distances.}
\label{fig:str1}
\end{center}
\end{figure}

In this section we investigate the lattice structures of (GTO)$_{m}$(STO)$_{n}$ superlattices for different $n$ and different electronic phases.  We begin with a brief discussion of the bulk structures. For SrTiO$_3$ we focus on the high temperature phase which is nearly cubic with a crystallographic unit cell containing one Ti ion and an octahedral volume of $\sim$ 9.9 $\AA^{3}$. (In bulk STO, a transition occurs at $T\sim110 K$ to a lower symmetry phase only slightly distorted from the high temperature structure.\cite{Unoki67} The differences between the high and low T phase  are not important for our discussion.) GdTiO$_{3}$ is strongly distorted away from the cubic perovskite structure, with a unit cell containing two Ti ions, a Ti-O-Ti bond angle $\approx 145^{\circ}$ and a substantially increased octahedral volume $\sim 11.3\;  \AA^{3}$. Further GdTiO$_3$ displays a high degree of $(\pi,\pi,0)$ ``orbital ordering'', involving a two sublattice pattern in which in a plane (which we take to be perpendicular to the growth direction) the Ti-O bonds in the x direction are elongated on one Ti sublattice and compressed on the other, with the bonds in the y direction behaving oppositely.  The differences in the octahedral rotations will be seen to have important consequences. To capture these we choose a $\sqrt{2}\times\sqrt{2}\times 2l$ ($l:$ integer) computational unit cell such that the two inequivalent Ti ions lie in the plane perpendicular to the growth direction.

Fig.~\ref{fig:str1} presents relaxed lattice structures obtained using a moderate $U=3$ eV so that the superlattice is metallic.   The top panel (a) gives atomic positions. The second panel (b) shows that the octahedral volume changes discontinuously across the interface, taking essentially the bulk GdTiO$_{3}$ value for Ti ions bounded on all sides by Gd ions and taking essentially the bulk SrTiO$_{3}$ value for Ti ions either bounded on all sides by Sr ions, or at the interfaces with two Gd and two Sr neighbors. The spatial variation of the structure is related to the band alignments, which are such that the charge density for all Ti ions with eight Gd neighbors takes essentially the bulk GdTiO$_{3}$ value. We will show below that the correlation effects expressed here by the U make an important contribution to the band alignments. The third panel (c) shows the variation of the Ti-O-Ti bond angle for both bonds in the planes ($\theta_{\parallel}$) and along the growth direction ($\theta_{\perp}$). We see again that the bond angle deviates significantly from the bulk GdTiO$_{3}$ value only in Ti layers surrounded by Sr or for ``apical'' oxygens connecting an interface Ti to a Sr.

\begin{figure}[htbp]
\begin{center}
\includegraphics[width=0.8\columnwidth, angle=-0]{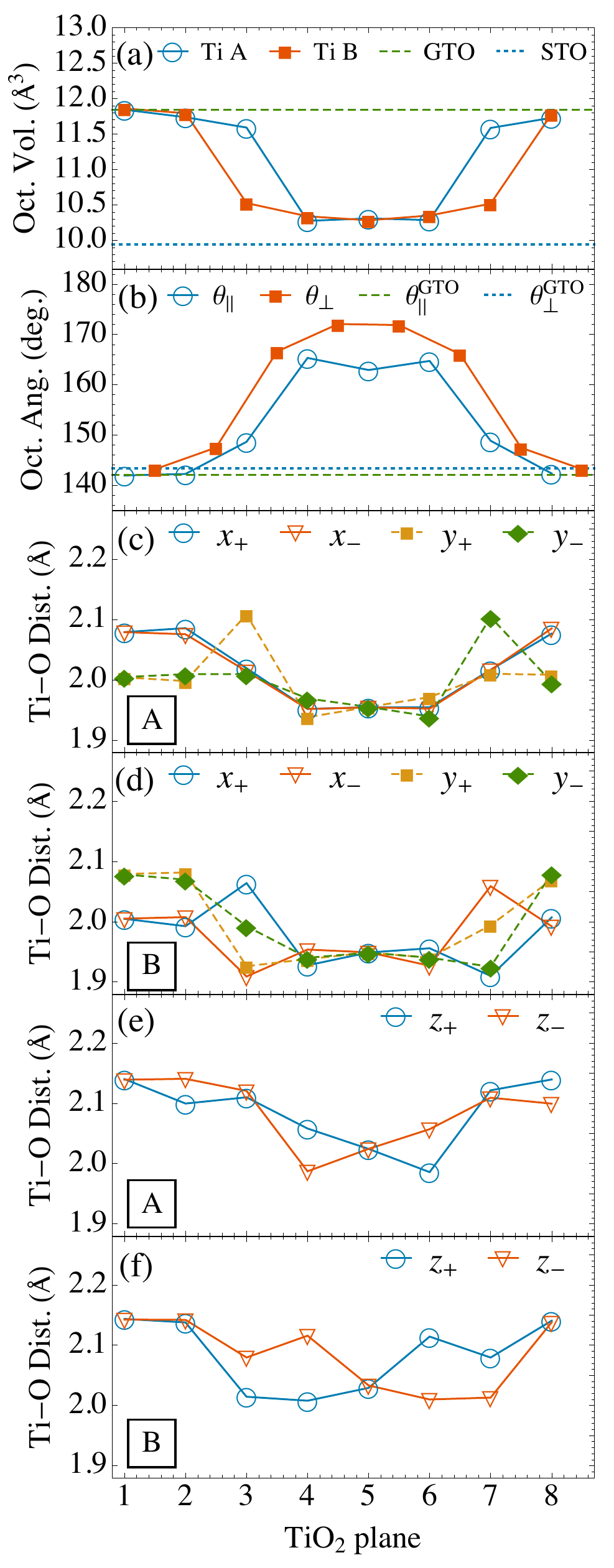}
\caption{(color online) Structural properties of (GTO)$_{4}$(STO)$_{4}$ superlattices in the insulating phase computed as described in the main text with $U=4$ eV. Atomic coordinates are same as shown in Fig.~\ref{fig:str1}. (a) Octahedral volumes around A and B sublattice Ti atoms. The green dashed and blue dotted lines are respectively the octahedral volume of bulk GTO and STO calculated with experimental in-plane lattice constant. (b) Average values of in-plane ($\theta_{||}$) and out-of-plane ($\theta_{\perp}$) Ti-O-Ti angles. The green dashed and blue dotted lines are the in-plane and out-of-plane octahedral angle for GTO calculated with experimental substrate, respectively. (c-f) Ti-O distance between A and B sublattice Ti atoms following the same definition in the Fig.~\ref{fig:str1}.}
\label{fig:str2}
\end{center}
\end{figure}

The remaining panels show Ti-O bond lengths. Bulk GdTiO$_{3}$ has a two-sublattice orbital ordering; our computational unit cell is such that the ordering wavevector is perpendicular to the growth direction (as is physically reasonable), and the resulting order is an in-plane $xz/yz$ alternation. Fig.~\ref{fig:str1} (d) shows the in-plane bond lengths for one of the two sublattices. We see that in the GdTiO$_{3}$ region (layers 1,2,8) the Ti-O bond length in the plus and minus $x$ directions is $\sim 0.1 \; \AA$ greater than the Ti-O bond length in the plus and minus $y$ directions (on the other sublattice, the pattern is reversed). Similarly within the SrTiO$_{3}$ (layers 4,5,6) there is negligible bond-length disproportionation. Thus as far as the in-plane bonds are concerned, the lattice distortions associated with GdTiO$_{3}$ propagate only as fas as the interface layer, consistent with what was observed for the bond angles and octahedral volumes. 

However, the in-plane bonds in the interface layer and the out-of-plane bonds throughout the SrTiO$_{3}$ region show an interesting behavior implying breaking of the local inversion symmetry at the Ti site. The inversion symmetry breaking in the $z$ direction may be understood in terms of the dielectric properties of SrTiO$_{3}$. The electric fields associated with the polar discontinuity of the GdTiO$_{3}$/SrTiO$_{3}$ interface and the associated induced charges are partly screened by polar modes of the SrTiO$_{3}$ lattice (in particular an off-centering of the Ti) as discussed in Okamoto {\it et al.}\cite{Okamoto06}. Our new finding is a difference in the $x$ and $y$ bonds implying a spontaneous ferroelectric distortion in the plane. We also note that there is no difference in the octahedral volumes between A and B sublattices implying equal electron occupancy of the A and B sublattice Ti states. We define this structure with the same octahedral volume for A and B sublattices as the non charge-ordered (NCO) phase.

In Fig.~\ref{fig:str2} we present the structural properties of insulating (GTO)$_{4}$(STO)$_{4}$ superlattices obtained using DFT+U with  $U=4$ eV. The variation of octahedral volume is very similar to that found in the previous metallic case, except that in the interface layer a clear difference between sublattices is visible. One sublattice (``A'') has almost the same octahedral volume as in bulk GdTiO$_{3}$ while the other has almost the same octahedral volume as bulk SrTiO$_{3}$. This difference reflects the almost complete charge order noted by many previous studies of related interfaces including LaAlO$_{3}$/SrTiO$_{3}$\cite{Pentcheva06} and LaTiO$_{3}$/SrTiO$_{3}.$\cite{Pentcheva07} The difference in octahedral volumes between sublattices has almost no effect on the in-plane bond angles; however the out of plane bond angels involving apical oxygens around interface Ti ions show small differences (about 3$^{\circ}$) between A- and B- sublattices (not shown here.) The magnitude of the centrosymmetry breaking distortions in the $z$-direction Ti-O bond is much less, reflecting the almost complete confinement of polar charge to the interface layer, strongly reducing the need to screen internal electric fields. We define this structure with inequivalent octahedral volumes as the charge order (CO) phase.

\begin{figure}[htbp]
\begin{center}
\includegraphics[width=0.8\columnwidth, angle=-0]{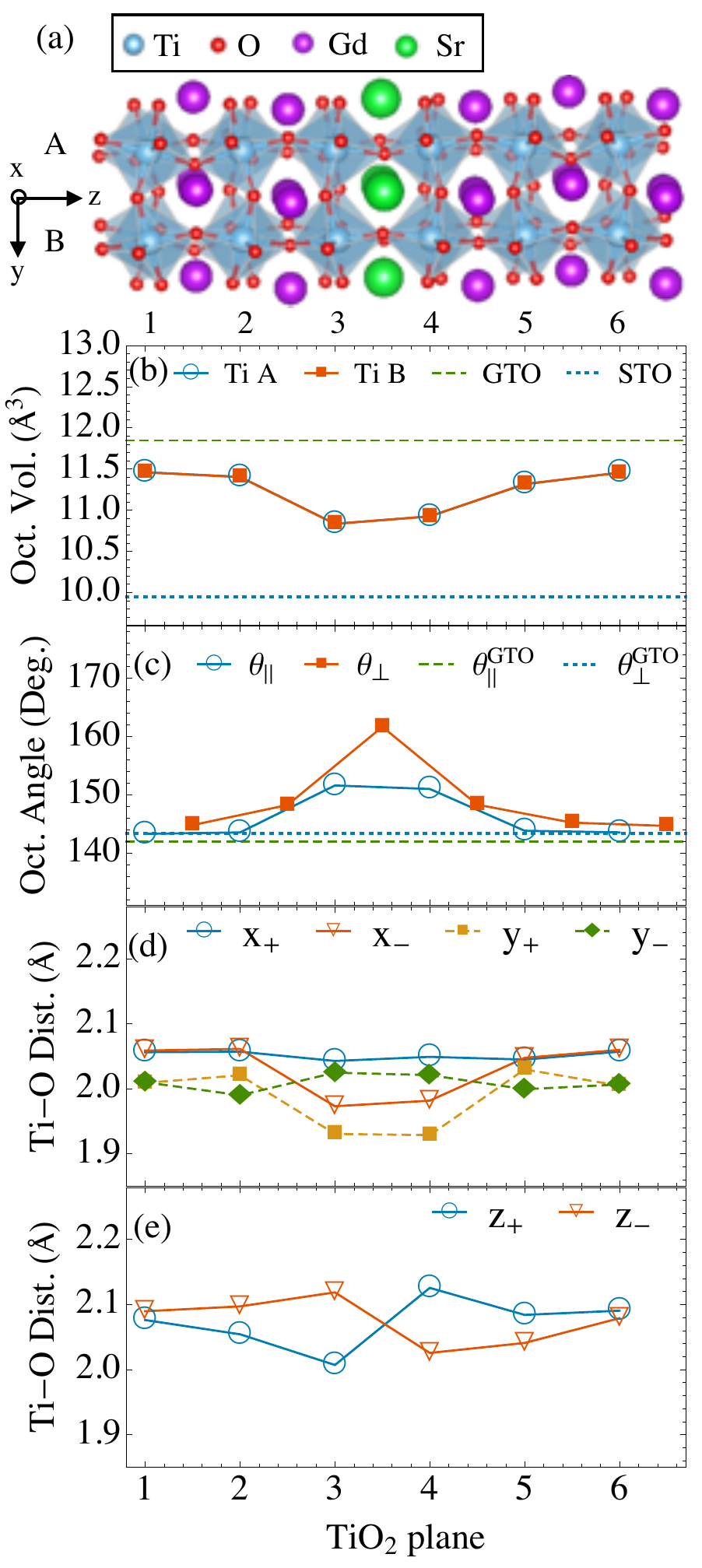}
\caption{(color online) Structural properties of (GTO)$_{5}$(STO)$_{1}$ superlattices computed as described in the main text with $U=2$ eV. (a) The atomic structure of the superlattice. A and B represent two different in-plane Ti atoms and the numbers in the bottom denote TiO$_{2}$ plane number.  (b) Octahedral volumes around A and B sublattice Ti atoms. The green dashed and blue dotted lines are respectively the octahedral volume of bulk GTO and STO calculated with the experimental lattice constant. (c) Average values of in-plane ($\theta_{||}$) and out-of-plane ($\theta_{\perp}$) Ti-O-Ti angles. The green dashed and blue dotted lines are the in-plane and out-of-plane octahedral angle for GTO calculated with experimental substrate, respectively. (d-e) Distance between Ti and O in the in-plane (d) and growth direction (e)  pseudo cubic direction of A sublattice Ti atoms. Exchanging $x_{\pm}$ to $y_{\mp}$ gives B sublattice Ti-O distances.}
\label{fig:str3}
\end{center}
\end{figure}

\begin{figure}[htbp]
\begin{center}
\includegraphics[width=0.8\columnwidth, angle=-0]{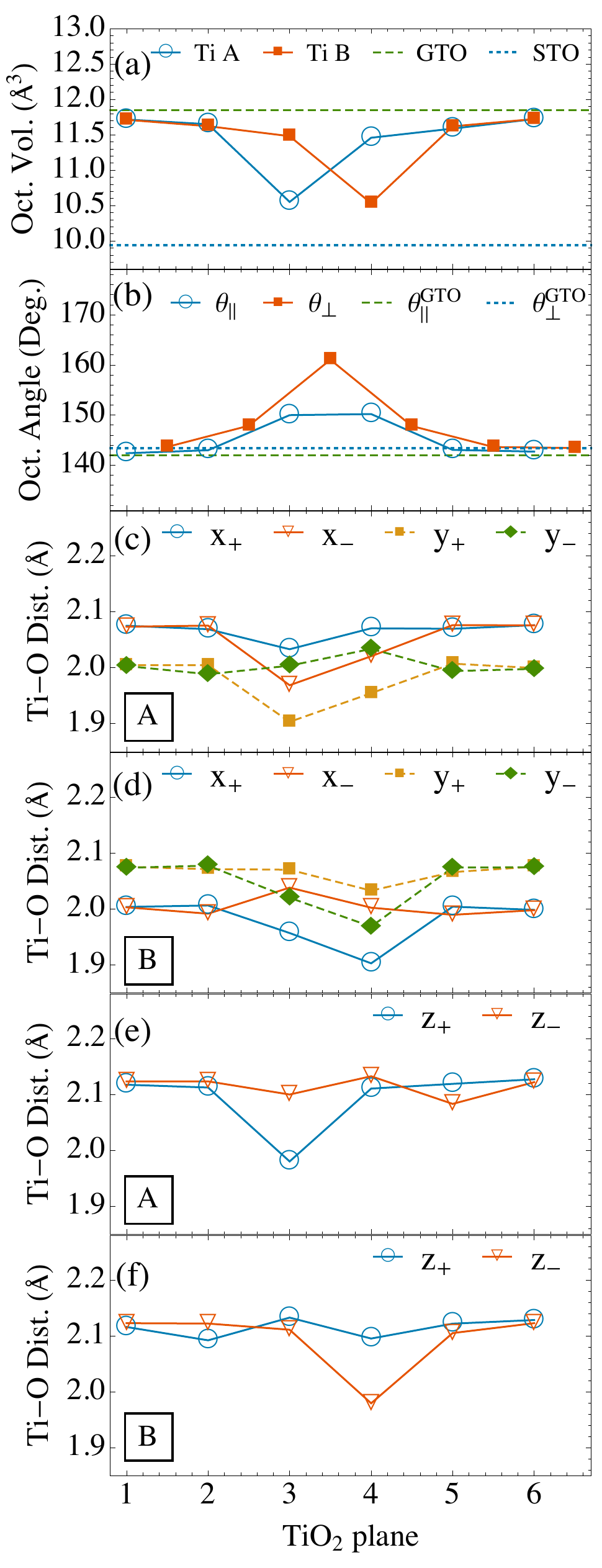}
\caption{(color online) Structural properties of (GTO)$_{5}$(STO)$_{1}$ superlattices in the insulating phase computed as described in the main text with $U=3$ eV. Atomic coordinates are same as shown in Fig.~\ref{fig:str3}. (a) Octahedral volumes around A and B sublattice Ti atoms. The green dashed and blue dotted lines are respectively the octahedral volume of GTO and STO calculated with the experimental lattice constant. (b) Average values of in-plane ($\theta_{||}$) and out-of-plane ($\theta_{\perp}$) Ti-O-Ti angles. The green dashed and blue dotted lines are the in-plane and out-of-plane octahedral angle for GTO calculated with experimental substrate, respectively. (c-f) Distance between Ti and O in each pseudo cubic direction of A and B sublattice Ti atoms.}
\label{fig:str4}
\end{center}
\end{figure}

For general (GTO)$_{m}$(STO)$_{n}$ superlattices, the structural transition from the NCO to the CO phase is driven by increasing the on-site Coulomb interaction $U$. However for $n=1$ superlattices with a single STO unit cell, the distortion around the interface is different when $U$ is small. Fig.~\ref{fig:str3} shows the structure for the metallic interface found in a $m=5$, $n=1$ superlattice with $U=2$ eV. As in the $n>1$ case there is no difference in the octahedral volume between A and B sublattice but there is a small difference in octahedral volume between layer 3 and 4 mainly due to the difference in the Ti-O distance in $z$ direction. This is different from the NCO phase of $n>1$ superlattices in which the octahedral volumes of two interfaces are different. We define the structure as CO II phase. We also note that the amount of Jahn-Teller distortion is much less in the GTO region due to the decreased correlation. As $U$ increases, the structure changes to sublattice charge order (CO) phase as presented in Fig.~\ref{fig:str4}. However unlike the $n>1$ case the preferred orbital character is not switched between GTO and the larger interface octahedron since the Ti-O distance in $x$ ($y$) direction are longer than $y$ ($x$) for all octahedra in A (B) sublattice.

\begin{figure}[htbp]
\begin{center}
\includegraphics[width=1\columnwidth, angle=-0]{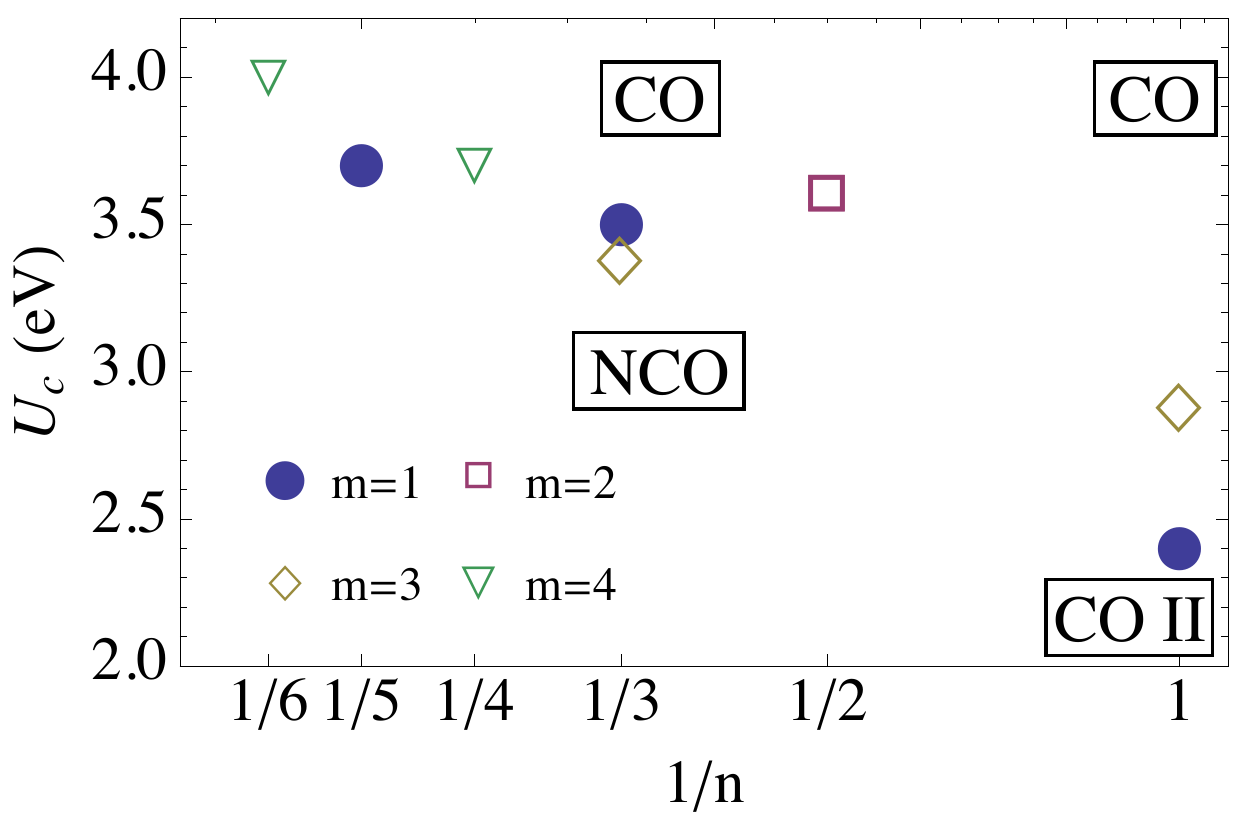}
\caption{(color online) Dependence of critical interaction strength $U_c$ for the structural phase transition on the number $n$ of STO layers in (GTO)$_{m}$(STO)$_{n}$ superlattices, computed as described in the text for different GTO thicknesses $m$ (shown with different symbols indicated in Figure legend). For $U<U_c$ the phase is either not charge ordered (NCO), for $n>1$ or weakly charge ordered (CO2) for $n=1$. For $U>U_c$ the phase is strongly charge ordered (CO) at all $n$.}
\label{fig:pt}
\end{center}
\end{figure}

Figure~\ref{fig:pt} presents the critical electronic correlation $U_{c}$ required to derive the structural phase transition between two distinct distortion patterns for (GTO)$_{m}$(STO)$_{n}$ superlattices. $U_{c}$ is defined as the value of $U$ in which the total energy with CO phase becomes lower than that with NCO (CO II) phase. The $U_{c}$ is relatively insensitive to the thickness of GTO and we can see a general trend that $U_{c}$ increases as the thickness of STO increases for $n>1$. However, there is drastic decrease in the $U_{c}$ for $n=1$ superlattices. Thus we can identify two different groups of $U_{c}$; one around 3.5 eV for $n > 1$ and around  2.5 eV for $n=1$. The difference in the critical U and structural transitions for $n=1$ and $n>1$ superlattices implies that the change in the underlying electronic structures between two groups of superlattices will be different.

\section{Electronic structure of $n>1$ GTO/STO superlattices}
\label{sec:twolayer}
This section presents the electronic structures of superlattices with two and more SrO layers. The CO distortion that we identified in the previous section plays an important role in localizing the electrons. The nominal valence counting implies that each GdO layer donates one-half  electron per in-plane unit cell to each adjacent TiO$_{2}$ layer. Thus, the interface TiO$_{2}$ layer is doped by an half electron per in-plane Ti. Therefore within band theory both charge and orbital ordering are required to obtain insulating states. Further, as shown in panels (c) and (f) of Fig.~\ref{fig:str2}, the apical oxygen is displaced in the $z$ direction, implying the $xy$ orbital is higher in energy than the $xz/yz$ orbitals. Thus, in this case the electrons are in two orbitally degenerate bands and insulating behavior requires charge, spin, and orbital order.

\begin{figure}[htbp]
\begin{center}
\includegraphics[width=1\columnwidth, angle=-0]{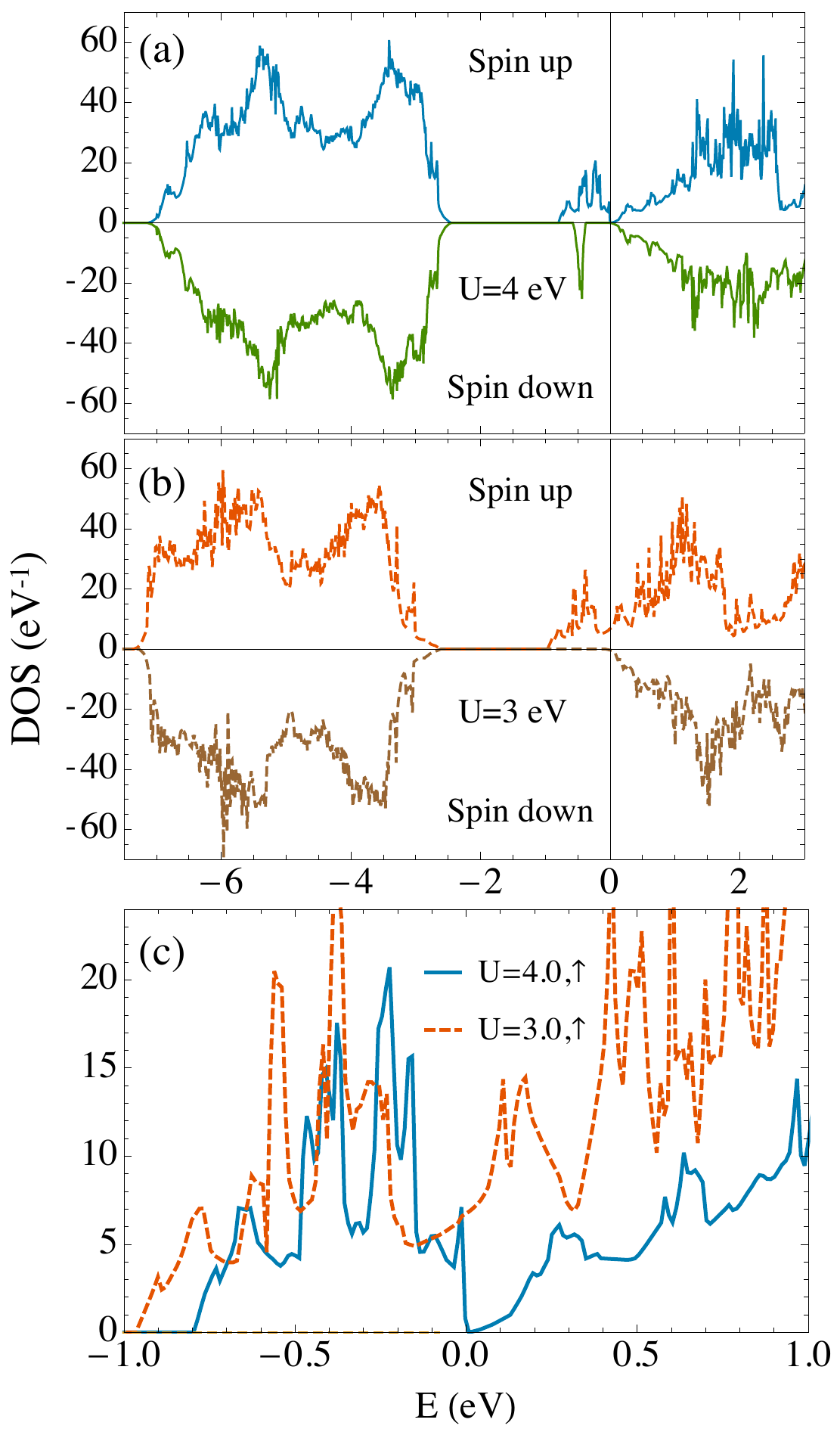}
\caption{(color online) Density of states of (GTO)$_{4}$(STO)$_{4}$ superlattices. (a) DOS for $U=4.0$ eV. (b) DOS for $U=3.0$ eV.   (c) Comparison of spin up density of states near the Fermi energy.}
\label{fig:dos}
\end{center}
\end{figure}

Fig.~\ref{fig:dos} shows the total density of states (DOS) of a (GTO)$_{4}$(STO)$_{4}$ heterostructure with $U=3$ (no charge order) and 4 eV (charge order). The Fermi energy is defined to be $E=0$. The states in between -7.5 and -2.5 eV are oxygen $p$ states and those near and above the Fermi energy are Ti-$t_{2g}$ derived states. The oxygen $p$-states shifts about 0.5 eV as the values of $U$ increases from 3 to 4 eV while the shift in energy of the Ti-$t_{2g}$ states is about 0.2 eV. There are evident changes in DOS near Fermi energy. For $U=4$ eV, the system is insulating  with small gap less than 0.1 eV. For $U=3$ eV, there are substantial increases in the density of states at the Fermi level and the system is a half-metal. As expected we find that the insulating ground state is associated with a volume difference in octahedra between interface  A-B sublattice, while in the metallic ground state all octahedra have equal volumes. 

\begin{figure*}[htbp]
\begin{center}
\includegraphics[width=2\columnwidth, angle=-0]{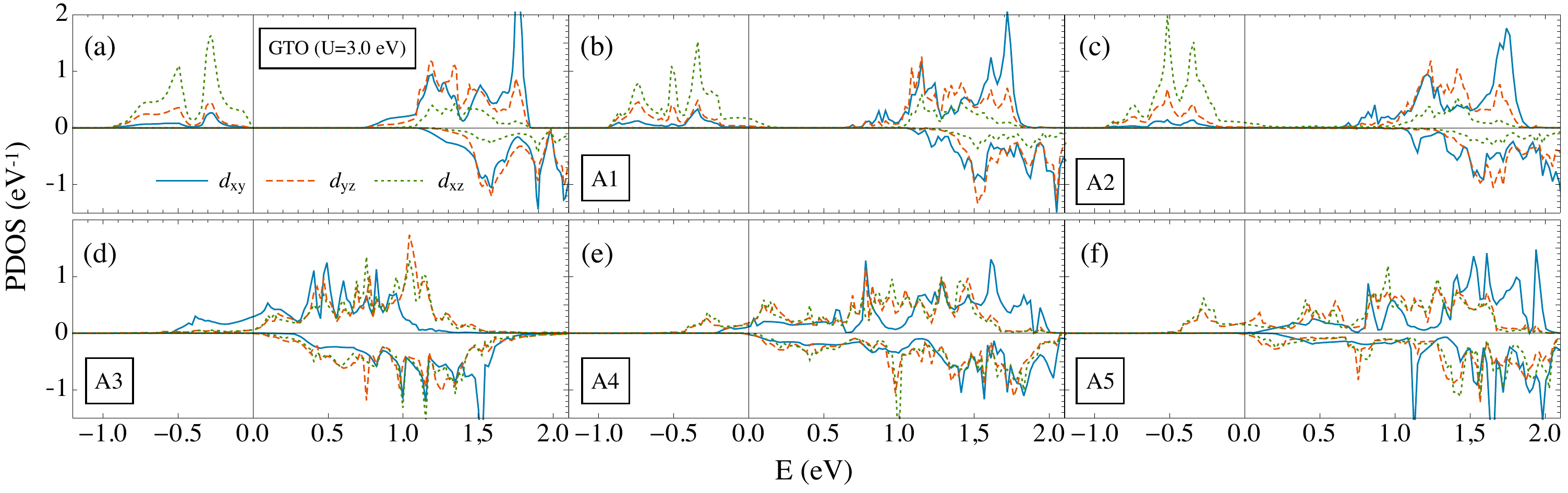}
\caption{(color online) Projected density of states (PDOS) of Ti-$t_{2g}$ orbitals. The projection is done with $t_{2g}$ orbital defined with pseudo cubic axes that deviate with octahedral coordinates but the change is not significant. Blue solid lines, red dashed lines, and green dotted lines represent $d_{xy}, d_{yz}$, and $d_{xz}$ orbitals, respectively. (a) PDOS of GTO with $U=$ 3 eV. (b-f) PDOS of (GTO)$_{4}$(STO)$_{4}$ with $U=$ 3 eV for A-sublattice Ti atoms - numbers indicate atoms defined in Fig. \ref{fig:str1}.}
\label{fig:pdosu3}
\end{center}
\end{figure*}

\begin{figure}[htbp]
\begin{center}
\includegraphics[width=0.9\columnwidth, angle=-0]{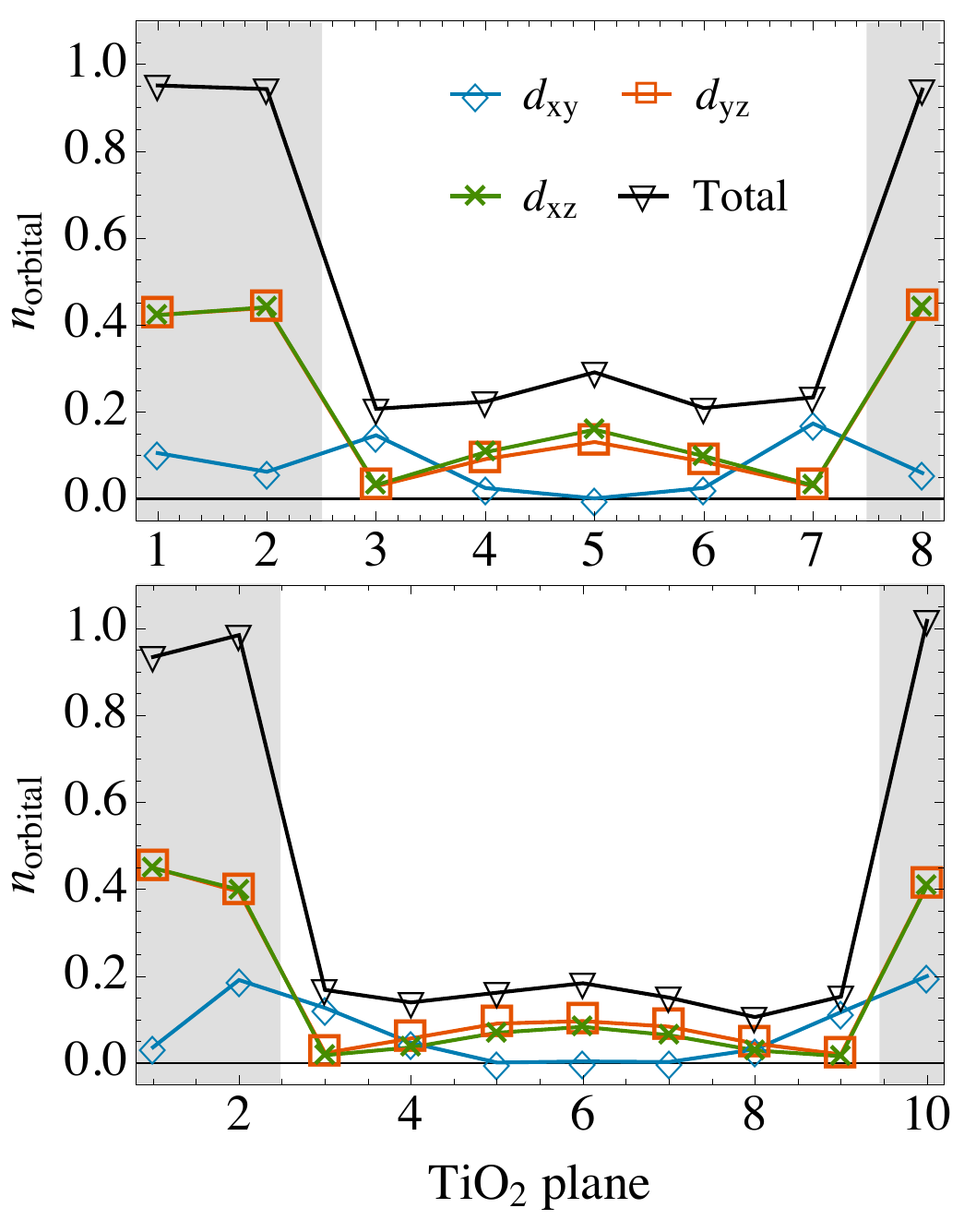}
\caption{(color online) Orbitally resolved and total occupancy of Ti-$t_{2g}$ states computed for $U=3$ eV (metallic regime) of (a) (GTO)$_{4}$(STO)$_{4}$ and (b)  (GTO)$_{4}$(STO)$_{6}$ superlattices. The gray region represents the Ti atoms between two GdO planes.}
\label{fig:cd}
\end{center}
\end{figure}

In order to investigate the nature of the conducting interface, we present the projected density of states (PDOS) of Ti-$t_{2g}$ derived bands for $U=3$ eV in Fig.~\ref{fig:pdosu3}.  For the A-sublattice Ti atoms in between GdO layers (panel (b) and (c) labeled as A1 and A2), the density of states is similar to that of bulk GTO (panel (a)) with negligible density of state at the Fermi level and is orbitally ordered ($xz$ on A-sublattice, $yz$ on B-sublattice). The sublattice orbital ordering disappears for Ti atoms at the interface and in between SrO layers. At the interface (panel (d)), the $xy$-derived band is dominantly occupied and away from the interface the occupancy of $yz/xz$ electrons is gradually increasing (panel (e-f)). These characteristic features of the PDOS are also found in other conducting superlattices with $n>1$. 

In Fig.~\ref{fig:cd}, we present the layer resolved occupation of the $t_{2g}$ orbitals of (GTO)$_{4}$(STO)$_{4}$ and (GTO)$_{4}$(STO)$_{6}$ superlattices showing that the orbital disproportionation persists to relatively thick superlattices. To define the occupation we integrate the PDOS of Ti $t_{2g}$ orbitals from -1 eV to Fermi energy and normalize the result so that the total integrated density of states equals two electrons. We find that for the relatively thick  STO layers studied the total $t_{2g}$ occupation does not vary much in the STO and interface region because the screening length is larger then the STO thickness. We believe that for both (GTO)$_{4}$(STO)$_{4}$ and (GTO)$_{4}$(STO)$_{6}$  superlattices this is an interface driven phenomenon similar to that found in LaAlO$_{3}$/SrTiO$_{3}$ interfaces, which have a dominantly occupied $xy$ orbital in the near-interface region and $yz/xz$ orbitals away from the interface\cite{Son209,Delugas11}.

\begin{figure*}[htbp]
\begin{center}
\includegraphics[width=2\columnwidth, angle=-0]{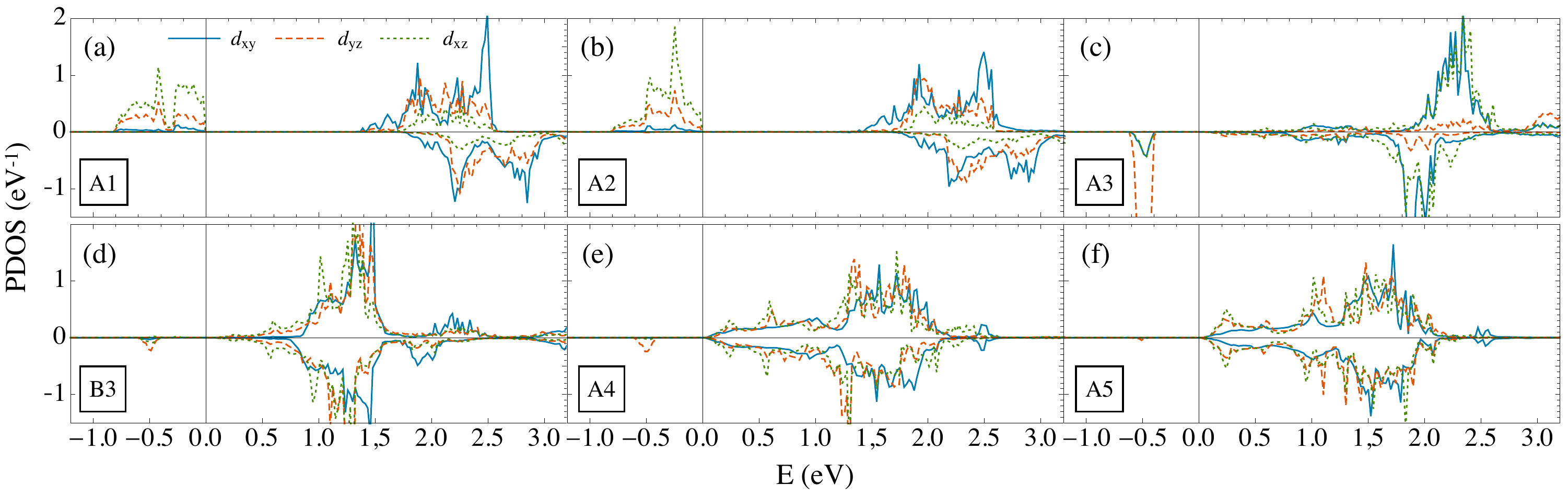}
\caption{(color online) (a-f) PDOS of Ti-$t_{2g}$ orbitals for (GTO)$_{4}$(STO)$_{4}$ superlattice with $U$=4 eV following the atomic index defined in Fig. \ref{fig:str1}. The label at the bottom left of each panel represents Ti atoms where the letter A and B denote sublattice index and numbers indicate atoms defined in Fig.~\ref{fig:str1}.  Blue solid lines, red dashed lines, and green dotted lines represent $d_{xy}, d_{yz}$, and $d_{xz}$ orbitals, respectively.}
\label{fig:pdosu4}
\end{center}
\end{figure*}

Fig.~\ref{fig:pdosu4} shows the PDOS of insulating ground state with $U=4$ eV. In this phase we find that the  electrons induced in the STO  are tied to interface Ti atoms and exhibit sublattice orbital and charge ordering. As in the conducting interfaces, the PDOS of Ti sandwiched by GdO layers (panel (a) and (b)) resembles that of bulk GTO. At the interface, the Ti states of the larger octahedron (A3, panel (c)) are occupied with magnetic moment 0.82 $\mu_{B}$ whereas the Ti states of smaller octahedron (B3, panel (d)) have negligible occupancy with magnetic moment 0.04 $\mu_{B}$. Compared with the magnetic moment of Ti in the middle of GTO (0.88 $\mu_{B}$), we can consider the charge ordering as nominally one and zero electron for larger and smaller octahedron, respectively. The occupied Ti atoms have ferro-orbital (in this case $yz$) and ferromagnetic order but with spin direction opposite to that in the GTO region. As pointed out in the previous section, the change in the orbital character from $xz$ to $yz$ along the A sublattice (panel (b) and (c)) is related to the change in the bond disproportionation between GTO region and interface octahedra. This in turn enhances the localization of interface Ti states due to the small transfer integral between different $t_{2g}$ orbitals. We note that the occupied states at the larger interface octahedron (panel (c)) have larger binding energy than the occupied states of GTO region (panel (a-b)) although the interface octahedral volume is slightly smaller than that of GTO region. Away from the interface the orbital and charge ordering disappears and the unoccupied $t_{2g}$ bands are shifted  closer to the Fermi energy. The PDOS for other (GTO)$_{m}$(STO)$_{n}$ interfaces with $n > 1$ are similar.

\section{Electronic structures with a single S\MakeLowercase{r}O layer}
\label{sec:onelayer}
In this section, we present the electronic structure of (GTO)$_{5}$(STO)$_{1}$ superlattices. In these superlattices  the critical $U$ for structural transition is significantly lower than in superlattices with multiple SrO layers. Unlike the previous cases with $n>1$, we will show that charge ordering is not a necessary condition for obtaining a insulating ground state and will show that the insulating phase persists to smaller $U$ value due to the lack of low lying unoccupied states. 
	
\begin{figure}[htbp]
\begin{center}
\includegraphics[width=0.8\columnwidth, angle=-0]{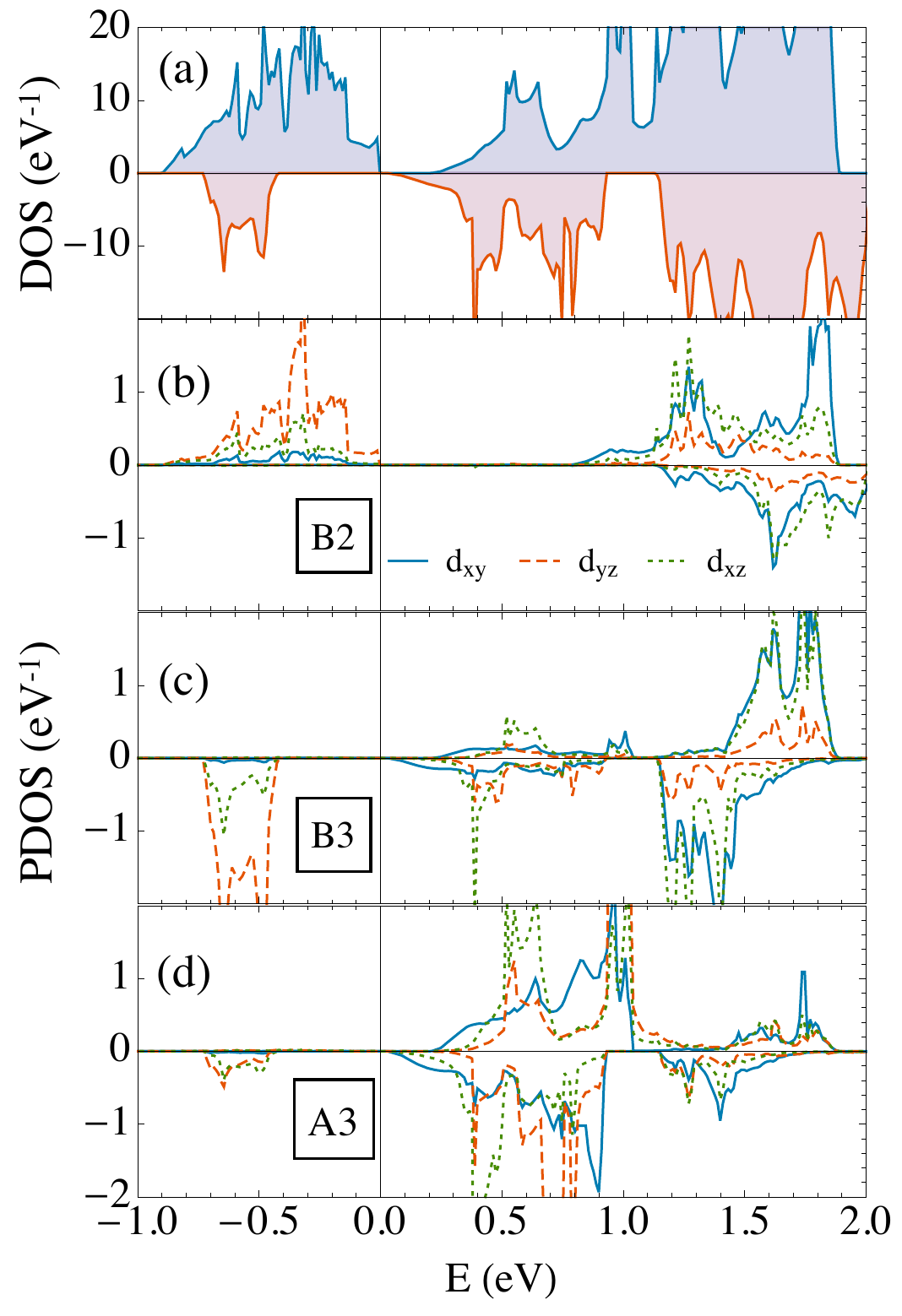}
\caption{(color online) Density of states of (GTO)$_{5}$(STO)$_{1}$ superlattices with $U=$ 3 eV. (a) DOS near the Fermi level. (b-d) Projected density of states of Ti-$t_{2g}$ orbitals for Ti positions defined in Fig.~\ref{fig:str3}. Blue solid lines, red dashed lines, and green dotted lines represent $d_{xy}, d_{yz}$, and $d_{xz}$ orbitals, respectively. }
\label{fig:dos2}
\end{center}
\end{figure}

The major difference between the electronic structures for single and multiple SrO layers is the absence, in the single-layer case, of the states coming from Ti atoms in between SrO layers. This significantly decreases the critical of $U$ for metal-insulator transition. Fig.~\ref{fig:dos2} presents the DOS and PDOS of (GTO)$_{5}$(STO)$_{1}$ superlattices with $U=3$ eV. From the total density of states we can see that the system is insulating with a band gap about 50 meV.  As seen in the PDOS, the Ti atom surrounded by GdO layers (panel (b)) has a density of states similar to that of bulk GTO. In our results with $U=3$ eV there exists both charge and orbital ordering but the difference in the magnetic moments of large (0.75 $\mu_{B}$) and small octahedra (0.14 $\mu_{B}$) is less than observed for $U=4$ eV. This indicates that in-plane charge ordering is not crucial to the insulating phase. This conclusion is supported by the following argument. With two TiO$_{2}$ layers, $yz\;(xz)$ orbitals in each layer form bonding and antibonding states with one dimensional dispersion in $y\; (x)$ direction. Due to this one-dimensional spectrum, if the electrons are ferromagnetically ordered, the band is half-filled with strong Fermi surface nesting. Given that the inter-plane hopping is not significantly smaller than intra-plane hopping and $xy$ band is higher in energy, a small orbital ordering that breaks the translation symmetry in both $x$ and $y$ direction and can open a gap at the Fermi energy. Thus as long as the octahedral distortion and electron correlation keep the $xy$ band unoccupied, the system becomes insulating even at weak correlation. We note that adding an additional STO layer will make the system metallic with $U=3$ eV since isolating a lowest $t_{2g}$ band is much harder because the low lying unoccupied band from the added STO layer will cross the Fermi energy.  In other words, lack of low lying unoccupied states and separation of bonding and antibonding bands makes it possible for $n=1$ superlattices insulating for lower $U$ value. Our results are consistent with the previously reported Hartree-Fock calculation by Chen \textit{et al.}\cite{Chen13} but for larger $U$ we find sublattice charge ordering in additional to the orbital ordering.

\section{Phase digram for structural and electronic phases transitions}
\label{sec:PT}

\begin{figure}[htbp]
\begin{center}
\includegraphics[width=1\columnwidth, angle=-0]{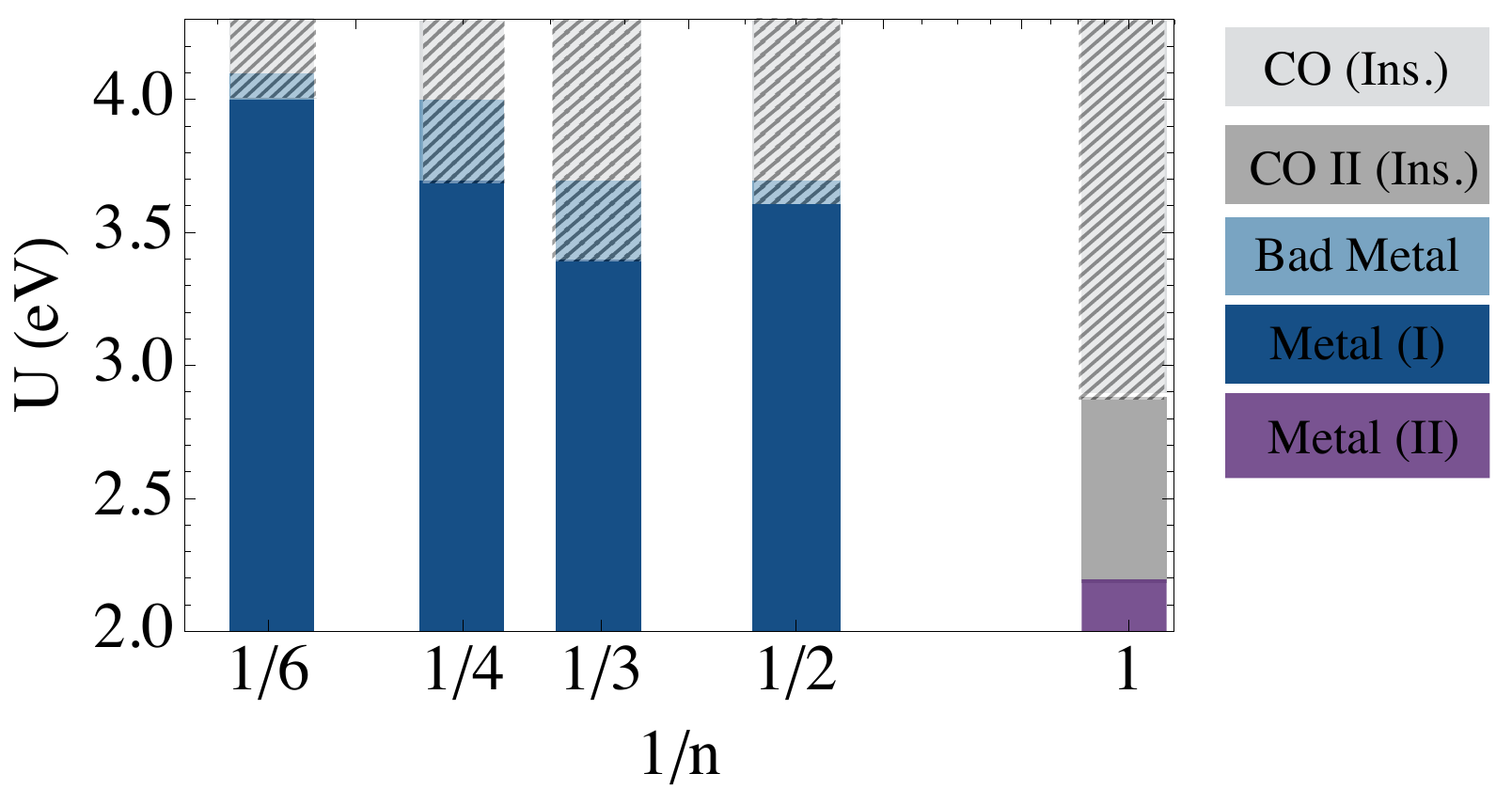}
\caption{(color online) Electronic and structural phase diagram for (GTO)$_{m}$(STO)$_{n}$ superlattices. We define in-plane and layered charge ordered insulating phase with orbital order as CO and CO II in gray and dark gray color, respectively. The structural CO phase is denoted as hatched lines so that the structural transition from CO phase to NCO (or CO II for $n=1$) is expressed as the boundary of the shaded region. For $n>1$, the narrow metallic region with charge ordering having very small density of state at the Fermi level is denoted as bad metal phase in light blue color. The ferromagnetic metal without charge ordering is denoted as metal phase in dark blue color and metallic phase with weak-layered charge ordering as metal II in purple. For the $n=1,3$ superlattices phase boundaries are obtained with $m=2$, the $n=2$ superlattice with $m=2$, and $n=4,6$ superlattices with $m=4$. The use of different $m$ values is related to a small increase in the MIT phase boundary of $n=2$ case.}
\label{fig:mit}
\end{center}
\end{figure}

In the section, we present a phase diagram summarizing our computed results for structural and electronic phases for (GTO)$_{m}$(STO)$_{n}$ superlattices. Fig.~\ref{fig:mit} shows the electronic and structural phase diagram in terms of critical $U$ values and inverse of the STO layer thickness $n$. We can see that for $n>1$, the MIT is accompanied by the structural transition which is crucial for the insulating phase. There is a narrow region of metallic phase before the structural transition with in-plane charge ordering for $n>1$ superlattices but the density of state is small at the Fermi energy. We identify this as a bad metal. 

Superlattices with a single SrO layer show qualitatively different phase behavior. Unlike the $n>1$ case, there is no sharp change in the electronic structure across the structural transition from CO to CO II phase and the MIT occurs maintaining a weak layered charge order. This supports the idea\cite{Chen13} that charge ordering is not necessary for $n=1$ superlattices. Moreover the value of the critical $U$ for the MIT of the $n=1$ superlattice is significantly smaller than $n>1$ superlattices which also supports the idea that weak correlation can open the band gap at the Fermi energy. Experimentally there is a MIT in between $n=3$ and $n=2$.\cite{Moetakef12} This is different from the prediction from the GGA+U phase diagram where the thickness dependent MIT occurs as the number of SrO layer changes from two to one.

\section{Broken inversion symmetry and ferroelectricity}
\label{sec:invsym}

As can be seen from the Ti displacements in Fig.~\ref{fig:str1}(d) or Fig.~\ref{fig:str2}(c-d), in appropriate circumstances the superlattices we consider can develop non-centrosymmetric distortions leading to in-plane ferroelectric polarization. In bulk GTO, the Gd ions move substantially off center ($\sim 0.5 \; \AA$) relative to the ideal perovskite position due to the the $a^{-}a^{-}c^{+}$ octahedral rotation but the displacements alternate from layer to layer in the (001) direction so the effects cancel and there is no net polarization, and in particular the Ti atoms remain at center of the surrounding oxygen octahedron. Replacing a plane of Gd by Sr breaks the translation symmetry so that the polarization no longer cancels as previously suggested.\cite{Benedek12} The symmetry breaking also leads to an off-centering of the Ti atoms, suggesting an additional ferroelectric component. This shift of Ti atoms suggests that there can be ferroelectric moment as in PbTiO$_{3}$.

\begin{figure}[htbp]
\begin{center}
\includegraphics[width=0.8\columnwidth, angle=-0]{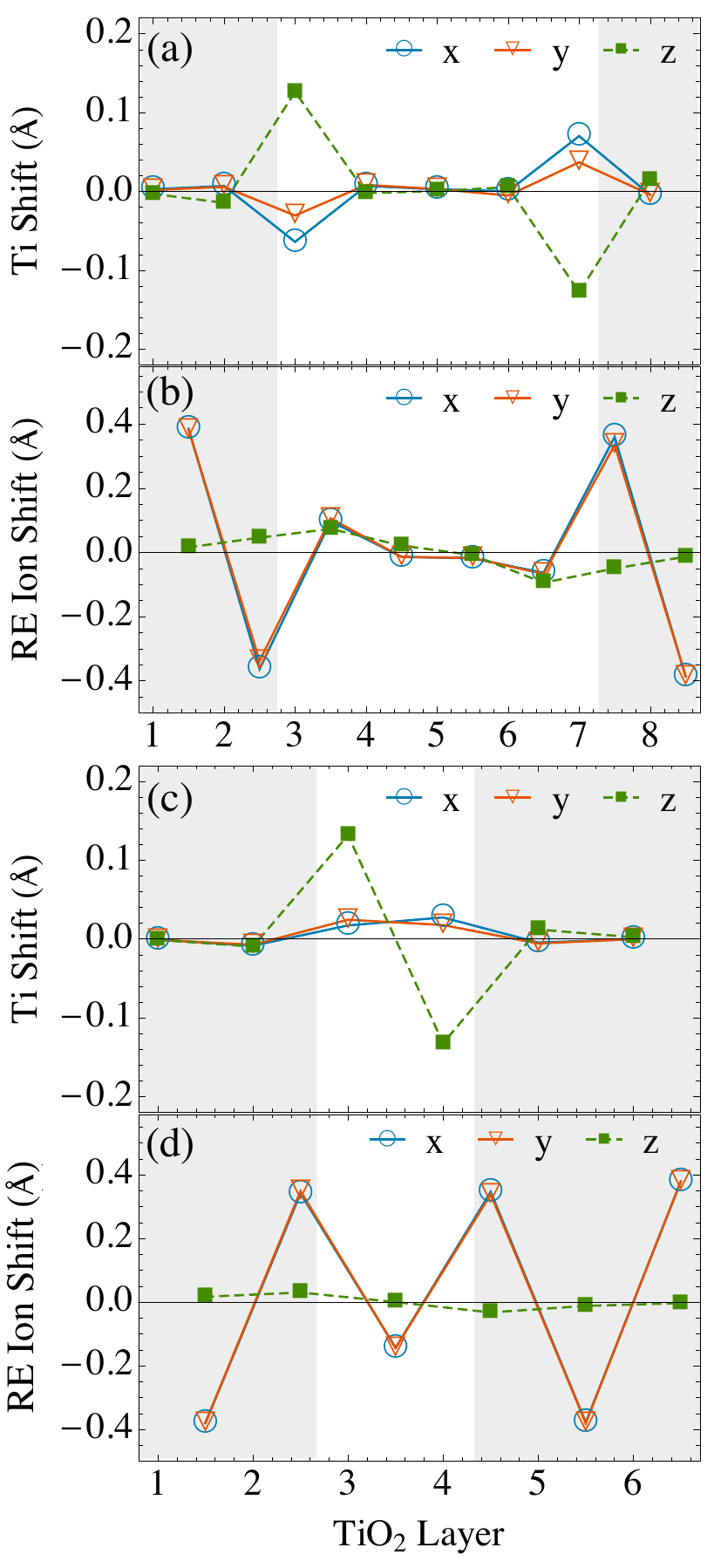}
\caption{(color online) The displacements of Ti and rare earth atoms (Gd, Sr) for insulating superlattices. (a-b) The displacements in (GTO)$_{4}$(STO)$_{4}$ superlattice with $U=4$ eV.  (c-d) The displacements in (GTO)$_{5}$(STO)$_{1}$ superlattice with $U=3$ eV. The displacements of Ti and rare earth atoms are defined as the distance from the center of mass of nearest neighbor in-plane oxygen atoms. The shaded regions represents the bulk-like GTO region. The difference in the displacements between A and B sublattice is small so averaged values are shown.}
\label{fig:epol}
\end{center}
\end{figure}

Fig.~\ref{fig:epol} shows atomic displacements of Ti and Gd atoms relative to the center of surrounding oxygen atoms for insulating (GTO)$_{4}$(STO)$_{4}$  ($U=4$ eV) and (GTO)$_{5}$(STO)$_{1}$ ($U=3$ eV) superlattices. In both superlattices, the shift is significant for near interface atoms. For out-of-plane direction, only interface Ti atoms are significantly displaced but the net moment is zero by the cancelation from the opposite contributions from two interfaces. For in-plane shifts, interface Ti atoms have the same displacement direction as that of Gd atoms next to the interface suggesting that the in-plane shift of Ti atoms is influenced by the movement of Gd atoms from the $a^{-}a^{-}c^{+}$ octahedral rotation. The magnitude of the in-plane shift is about 0.07 $\AA$ for Ti and 0.5 $\AA$ for Gd with (GTO)$_{4}$(STO)$_{4}$ superlattice and  0.03 $\AA$ for Ti and 0.5 $\AA$ for Gd atom with (GTO)$_{5}$(STO)$_{1}$ superlattice. For comparison, in PbTiO$_{3}$ the Ti and Pb atoms shift against the oxygen atoms about 0.26 $\AA$ and 0.42 $\AA$, respectively.\cite{Meyer99} Therefore we can see that the ionic shift is dominated by rare-earth atomic displacements.  The amplitude of rare earth ion motion in our superlattices is comparable to that found in the canonical ferroelectric PbTiO$_{3}$, but the amplitude of Ti atom motion in the superlattice is a about a factor of 5 smaller than the corresponding motion in PbTiO$_3$. There are also differences in ferroelectric polarization between the superlattices with even and odd number of STO layers. When there is an even number of STO layers, the net ferroelectric moment is zero due to the opposite polarization direction of two interfaces. Whereas with an odd number of STO layers, there is an in-phase shift from interface Ti atoms giving a small upward polarization and there is relatively larger upward polarization from the displacements of rare earth atoms. As seen in Fig.~\ref{fig:epol} (d) the uncompensated displacement of Gd atom located between the sixth and the first TiO$_{2}$ layer is larger than the displacement of Sr atom in between third and forth TiO$_{2}$ layers. Combined with the difference in nominal ionic charge between Gd (3+) and Sr (2+), there is a net upward polarization with an additional contribution from a small displacement of Ti atoms in layer 3 and 4 (Fig.~\ref{fig:epol} (c)). Therefore we expect that superlattices with an odd number of SrO layer, or more generally superlattices having two near interface Gd atoms displaced in the same direction may have a net ferroelectric moment. In the calculations performed so far, the polarization lies in the $xy$ plane, but `tricolor' superlattices with Gd-Sr-X may also have a $z$-direction polarization.

\section{Band alignment and metal-insulator transition}
\label{sec:ba}

\begin{figure}[htbp]
\begin{center}
\includegraphics[width=1\columnwidth, angle=-0]{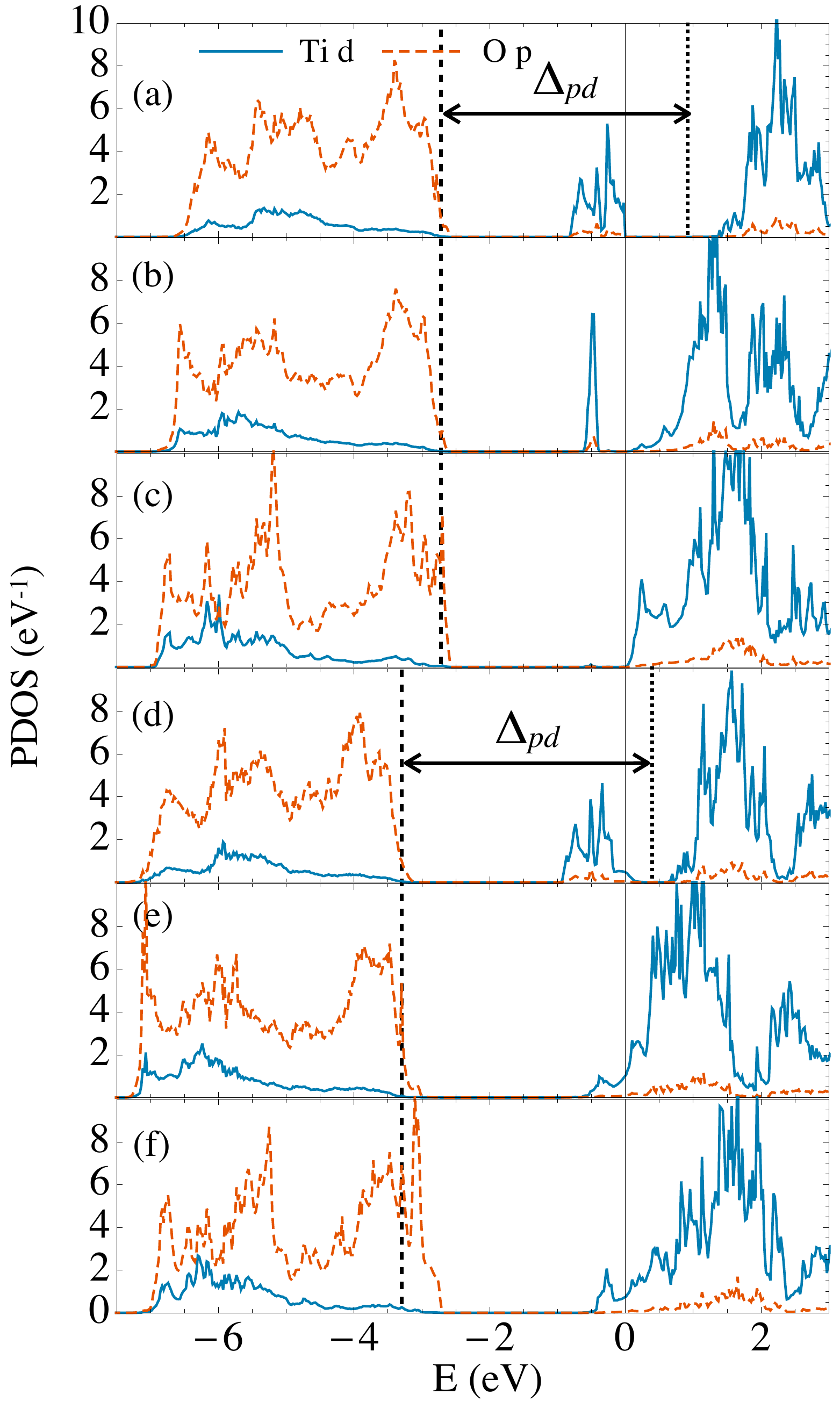}
\caption{(color online) PDOS of Ti $d$ (blue solid line) and oxygen $p$ (dotted red line) orbitals of (GTO)$_{4}$(STO)$_{4}$ superlattice.   (a) PDOS in the middle of GTO with  $U=3$ eV and (b) $U=4$ eV. (c) PDOS of interface TiO$_{2}$ layer for $U=3$ eV and (d) U=4 eV. (e) PDOS in the middle of STO for U=3 eV and (f) U=4 eV. The solid black line is the Fermi energy and the dashed black lines in the panel (a) and (b) represent the band edge of oxygen $p$ orbital in the middle of GTO. The energy difference between the dashed black line and dotted black line (center of gap between lower and upper Hubbard bands) is defined as $\Delta_{pd}$. In panel (e) and (f) the $E_{pol}$ is defined as the energy difference of oxygen band edge of GTO and STO region.}
\label{fig:bandalign}
\end{center}
\end{figure}

In the polar catastrophe scenario of LAO/STO interfaces, a MIT occurs as the number of LAO layers increases.\cite{Thiel06} The MIT is caused by a change in the band alignment of the STO conduction band driven by an increase in the electrostatic potential proportional to the LAO thickness.\cite{Nakagawa06,Millis10,Chen10,Son209} Similarly we can think of the metal insulator transitions of $n>1$ superlattices in terms of band alignments between the filled lower Hubbard band of GTO and the low-lying conduction band of STO. Fig.~\ref{fig:bandalign} shows the PDOS of oxygen $p$ and Ti $d$ orbitals in the bulk-like GTO region, at the interface, and bulk-like STO region. For the insulating interface, we can see that the lower Hubbard band of GTO in the bulk-like region lies slightly lower than the band bottom of conduction band in the bulk-like STO region. On the other hand, for metallic interfaces, the conduction band of the bulk-like STO region overlaps the occupied lower Hubbard band of the bulk-like GTO region. The band overlap is determined by four factors: the band gap in the GTO region $\Delta_{g}$,  $\Delta_{pd}$ defined as the energy difference between oxygen $p$ bands and the middle of upper and lower Hubbard bands, $\Delta_{STO}$ denoting the band gap of STO, and $E_{pol}$ defined as the energy shift in the top of oxygen $p$ bands between GTO and STO region. We note that the energy shift $E_{pol}$ arises from the dipole formed between GTO and STO. In the insulating phase, the energy shift is $\mathcal{E} d/2 \sim 2\pi e^{2} n_{2D} d / \epsilon$ where $\mathcal{E}$ is electric field between interface TiO$_{2}$ and adjacent GdO layer, $d$ is Ti-Ti distance, $\epsilon$ is a dielectric constant, and $n_{2D}$ is the sheet charge density of half electron per in-plane unit cell. In the metallic phase, electrons are delocalized over several layers of the STO region so the electric charge is farther from the interface and a larger dipole moment is expected. The energy shift is $\frac{4\pi e^{2} n_{2D}}{\epsilon} \int^{L}_{0} dz \mathcal{E}(z)  $ where $L$ the distance from the GdO layer next to the interface to the TiO$_{2}$ plane at the center of STO. The estimated energy shift for insulating superlattice with four STO layers is about 155 meV by substituting the sheet charge density of 0.5 electron per in-plane unit cell and dielectric constant 75 \cite{SP} and 388 meV for metallic superlattice by assuming uniformly distributed electron gas (as in Fig.~\ref{fig:cd}). These values are consistent with the energy shift in Fig.~\ref{fig:bandalign} which is about 200 meV in the insulating superlattice and 700 meV for metallic superlattice. The factor of two differences may come from the uncertainty in the dielectric constant.  

Thus, we can write the condition for obtaining an insulating superlattice as follows: 
\begin{eqnarray*}
E_{pol} + \Delta_{STO}(n) > \Delta_{pd}- \Delta_{g}(U)/2~.
\end{eqnarray*}
Given that $\Delta_{pd}$ and $E_{pol}$ is not sensitive to the value of $U$ and $n$, we can see  that the critical $U$ for MIT increases as the thickness of STO increases since  $\Delta_{STO}$ decreases as the number of STO layers $n$ increases due to the confinement effect,\cite{Lee13} while $\Delta_{g}$ is proportional to $U$. Although we neglect the energy cost of octahedral distortion and energy gain from localizing electrons, this relation gives a reasonable estimate of the ground state of heterostructures between band and Mott insulator with bulk properties the materials.

\section{Summary}
\label{sec:sum}
We investigated the metal-insulator transition of GTO/STO superlatives using first-principles GGA+U method.  Two different mechanisms for the insulating phase are identified. We find that charge and orbital ordering accompanied by difference in the volume of interface octahedra (``two sublattice charge order'') are necessary to obtain the insulating phase for superlattices with $n>1$. On the other hand,  in superlattices with $n=1$ the insulating gap emerges via a combination of orbital ordering (leading to a quasi-one dimensional in-plane dispersion) combined with a bonding-andtibonding splitting arising from coupling across the SrO layer, as previously found by  Chen {\it et al.}\cite{Chen13}. We find that the critical $U$ needed to drive the MIT for $n=1$ superlattices is significantly smaller than $n>1$ since small orbital ordering can open a gap in the bonding band due to the instability in the one dimensional dispersion of $yz/xz$ bands. A local inversion symmetry breaking around Ti atoms is observed and it is shown that ferroelectric polarization is possible with odd number of STO layers. We present the phase diagram for general (GTO)$_{m}$/(STO)$_{n}$  and within GGA+U showing discrepancy between our results and the transport measurement of Ref. \onlinecite{Moetakef12}, which shows a metal insulator transition already at $n=2$. The difference cannot be resolved by fine-tuning $U$ implying the importance of additional physics not included in the DFT+U approximation.

\section*{ACKNOWLEDGEMENTS}
This work is supported by DOE ER-046169 and Kwanjeong educational foundation. Computational facilities are provided via XSEDE resources through Grant No. TG-PHY130003.

\bibliography{park_manuscript.bib}

\begin{thebibliography}{44}
\expandafter\ifx\csname natexlab\endcsname\relax\def\natexlab#1{#1}\fi
\expandafter\ifx\csname bibnamefont\endcsname\relax
  \def\bibnamefont#1{#1}\fi
\expandafter\ifx\csname bibfnamefont\endcsname\relax
  \def\bibfnamefont#1{#1}\fi
\expandafter\ifx\csname citenamefont\endcsname\relax
  \def\citenamefont#1{#1}\fi
\expandafter\ifx\csname url\endcsname\relax
  \def\url#1{\texttt{#1}}\fi
\expandafter\ifx\csname urlprefix\endcsname\relax\def\urlprefix{URL }\fi
\providecommand{\bibinfo}[2]{#2}
\providecommand{\eprint}[2][]{\url{#2}}

\bibitem[{\citenamefont{Imada et~al.}(1998)\citenamefont{Imada, Fujimori, and
  Tokura}}]{Imada98}
\bibinfo{author}{\bibfnamefont{M.}~\bibnamefont{Imada}},
  \bibinfo{author}{\bibfnamefont{A.}~\bibnamefont{Fujimori}}, \bibnamefont{and}
  \bibinfo{author}{\bibfnamefont{Y.}~\bibnamefont{Tokura}},
  \bibinfo{journal}{Reviews of Modern Physics} \textbf{\bibinfo{volume}{70}},
  \bibinfo{pages}{1039} (\bibinfo{year}{1998}).

\bibitem[{\citenamefont{Ohtomo et~al.}(2002)\citenamefont{Ohtomo, Muller,
  Grazul, and Hwang}}]{Ohtomo02}
\bibinfo{author}{\bibfnamefont{A.}~\bibnamefont{Ohtomo}},
  \bibinfo{author}{\bibfnamefont{D.~A.} \bibnamefont{Muller}},
  \bibinfo{author}{\bibfnamefont{J.~L.} \bibnamefont{Grazul}},
  \bibnamefont{and} \bibinfo{author}{\bibfnamefont{H.~Y.} \bibnamefont{Hwang}},
  \bibinfo{journal}{Nature} \textbf{\bibinfo{volume}{419}},
  \bibinfo{pages}{378} (\bibinfo{year}{2002}).

\bibitem[{\citenamefont{Ohtomo and Hwang}(2004)}]{Ohtomo04}
\bibinfo{author}{\bibfnamefont{A.}~\bibnamefont{Ohtomo}} \bibnamefont{and}
  \bibinfo{author}{\bibfnamefont{H.~Y.} \bibnamefont{Hwang}},
  \bibinfo{journal}{Nature} \textbf{\bibinfo{volume}{427}},
  \bibinfo{pages}{423} (\bibinfo{year}{2004}).

\bibitem[{\citenamefont{Moetakef
  et~al.}(2011{\natexlab{a}})\citenamefont{Moetakef, Zhang, Kozhanov, Jalan,
  Seshadri, Allen, and Stemmer}}]{Moetakef11}
\bibinfo{author}{\bibfnamefont{P.}~\bibnamefont{Moetakef}},
  \bibinfo{author}{\bibfnamefont{J.~Y.} \bibnamefont{Zhang}},
  \bibinfo{author}{\bibfnamefont{A.}~\bibnamefont{Kozhanov}},
  \bibinfo{author}{\bibfnamefont{B.}~\bibnamefont{Jalan}},
  \bibinfo{author}{\bibfnamefont{R.}~\bibnamefont{Seshadri}},
  \bibinfo{author}{\bibfnamefont{S.~J.} \bibnamefont{Allen}}, \bibnamefont{and}
  \bibinfo{author}{\bibfnamefont{S.}~\bibnamefont{Stemmer}},
  \bibinfo{journal}{Applied Physics Letters} \textbf{\bibinfo{volume}{98}}
  (\bibinfo{year}{2011}{\natexlab{a}}).

\bibitem[{\citenamefont{Moetakef
  et~al.}(2011{\natexlab{b}})\citenamefont{Moetakef, Cain, Ouellette, Zhang,
  Klenov, Janotti, Van~de Walle, Rajan, Allen, and Stemmer}}]{Moetakef11-2}
\bibinfo{author}{\bibfnamefont{P.}~\bibnamefont{Moetakef}},
  \bibinfo{author}{\bibfnamefont{T.~A.} \bibnamefont{Cain}},
  \bibinfo{author}{\bibfnamefont{D.~G.} \bibnamefont{Ouellette}},
  \bibinfo{author}{\bibfnamefont{J.~Y.} \bibnamefont{Zhang}},
  \bibinfo{author}{\bibfnamefont{D.~O.} \bibnamefont{Klenov}},
  \bibinfo{author}{\bibfnamefont{A.}~\bibnamefont{Janotti}},
  \bibinfo{author}{\bibfnamefont{C.~G.} \bibnamefont{Van~de Walle}},
  \bibinfo{author}{\bibfnamefont{S.}~\bibnamefont{Rajan}},
  \bibinfo{author}{\bibfnamefont{S.~J.} \bibnamefont{Allen}}, \bibnamefont{and}
  \bibinfo{author}{\bibfnamefont{S.}~\bibnamefont{Stemmer}},
  \bibinfo{journal}{Applied Physics Letters} \textbf{\bibinfo{volume}{99}},
  \bibinfo{pages}{232116} (\bibinfo{year}{2011}{\natexlab{b}}).

\bibitem[{\citenamefont{Biscaras et~al.}(2010)\citenamefont{Biscaras, Bergeal,
  Kushwaha, Wolf, Rastogi, Budhani, and Lesueur}}]{Biscaras10}
\bibinfo{author}{\bibfnamefont{J.}~\bibnamefont{Biscaras}},
  \bibinfo{author}{\bibfnamefont{N.}~\bibnamefont{Bergeal}},
  \bibinfo{author}{\bibfnamefont{A.}~\bibnamefont{Kushwaha}},
  \bibinfo{author}{\bibfnamefont{T.}~\bibnamefont{Wolf}},
  \bibinfo{author}{\bibfnamefont{A.}~\bibnamefont{Rastogi}},
  \bibinfo{author}{\bibfnamefont{R.~C.} \bibnamefont{Budhani}},
  \bibnamefont{and} \bibinfo{author}{\bibfnamefont{J.}~\bibnamefont{Lesueur}},
  \bibinfo{journal}{Nat Commun} \textbf{\bibinfo{volume}{1}},
  \bibinfo{pages}{89} (\bibinfo{year}{2010}).

\bibitem[{\citenamefont{Seo et~al.}(2007)\citenamefont{Seo, Choi, Lee, Yu, Kim,
  Bernhard, and Noh}}]{Seo07}
\bibinfo{author}{\bibfnamefont{S.~S.~A.} \bibnamefont{Seo}},
  \bibinfo{author}{\bibfnamefont{W.~S.} \bibnamefont{Choi}},
  \bibinfo{author}{\bibfnamefont{H.~N.} \bibnamefont{Lee}},
  \bibinfo{author}{\bibfnamefont{L.}~\bibnamefont{Yu}},
  \bibinfo{author}{\bibfnamefont{K.~W.} \bibnamefont{Kim}},
  \bibinfo{author}{\bibfnamefont{C.}~\bibnamefont{Bernhard}}, \bibnamefont{and}
  \bibinfo{author}{\bibfnamefont{T.~W.} \bibnamefont{Noh}},
  \bibinfo{journal}{Phys. Rev. Lett.} \textbf{\bibinfo{volume}{99}},
  \bibinfo{pages}{266801} (\bibinfo{year}{2007}).

\bibitem[{\citenamefont{Ahn et~al.}(2003)\citenamefont{Ahn, Triscone, and
  Mannhart}}]{Ahn03}
\bibinfo{author}{\bibfnamefont{C.~H.} \bibnamefont{Ahn}},
  \bibinfo{author}{\bibfnamefont{J.~M.} \bibnamefont{Triscone}},
  \bibnamefont{and} \bibinfo{author}{\bibfnamefont{J.}~\bibnamefont{Mannhart}},
  \bibinfo{journal}{Nature} \textbf{\bibinfo{volume}{424}},
  \bibinfo{pages}{1015} (\bibinfo{year}{2003}).

\bibitem[{\citenamefont{Hwang et~al.}(2012)\citenamefont{Hwang, Iwasa,
  Kawasaki, Keimer, Nagaosa, and Tokura}}]{Hwang12}
\bibinfo{author}{\bibfnamefont{H.~Y.} \bibnamefont{Hwang}},
  \bibinfo{author}{\bibfnamefont{Y.}~\bibnamefont{Iwasa}},
  \bibinfo{author}{\bibfnamefont{M.}~\bibnamefont{Kawasaki}},
  \bibinfo{author}{\bibfnamefont{B.}~\bibnamefont{Keimer}},
  \bibinfo{author}{\bibfnamefont{N.}~\bibnamefont{Nagaosa}}, \bibnamefont{and}
  \bibinfo{author}{\bibfnamefont{Y.}~\bibnamefont{Tokura}},
  \bibinfo{journal}{Nat Mater} \textbf{\bibinfo{volume}{11}},
  \bibinfo{pages}{103} (\bibinfo{year}{2012}).

\bibitem[{\citenamefont{Zubko et~al.}(2011)\citenamefont{Zubko, Gariglio,
  Gabay, Ghosez, and Triscone}}]{Zubko11}
\bibinfo{author}{\bibfnamefont{P.}~\bibnamefont{Zubko}},
  \bibinfo{author}{\bibfnamefont{S.}~\bibnamefont{Gariglio}},
  \bibinfo{author}{\bibfnamefont{M.}~\bibnamefont{Gabay}},
  \bibinfo{author}{\bibfnamefont{P.}~\bibnamefont{Ghosez}}, \bibnamefont{and}
  \bibinfo{author}{\bibfnamefont{J.~M.} \bibnamefont{Triscone}}, in
  \emph{\bibinfo{booktitle}{Annual Review of Condensed Matter Physics, Vol 2}},
  edited by \bibinfo{editor}{\bibfnamefont{J.~S.} \bibnamefont{Langer}}
  (\bibinfo{publisher}{Annual Reviews}, \bibinfo{address}{Palo Alto},
  \bibinfo{year}{2011}), vol.~\bibinfo{volume}{2} of
  \emph{\bibinfo{series}{Annual Review of Condensed Matter Physics}}, pp.
  \bibinfo{pages}{141--165}.

\bibitem[{\citenamefont{Caviglia et~al.}(2008)\citenamefont{Caviglia, Gariglio,
  Reyren, Jaccard, Schneider, Gabay, Thiel, Hammerl, Mannhart, and
  Triscone}}]{Caviglia08}
\bibinfo{author}{\bibfnamefont{A.~D.} \bibnamefont{Caviglia}},
  \bibinfo{author}{\bibfnamefont{S.}~\bibnamefont{Gariglio}},
  \bibinfo{author}{\bibfnamefont{N.}~\bibnamefont{Reyren}},
  \bibinfo{author}{\bibfnamefont{D.}~\bibnamefont{Jaccard}},
  \bibinfo{author}{\bibfnamefont{T.}~\bibnamefont{Schneider}},
  \bibinfo{author}{\bibfnamefont{M.}~\bibnamefont{Gabay}},
  \bibinfo{author}{\bibfnamefont{S.}~\bibnamefont{Thiel}},
  \bibinfo{author}{\bibfnamefont{G.}~\bibnamefont{Hammerl}},
  \bibinfo{author}{\bibfnamefont{J.}~\bibnamefont{Mannhart}}, \bibnamefont{and}
  \bibinfo{author}{\bibfnamefont{J.~M.} \bibnamefont{Triscone}},
  \bibinfo{journal}{Nature} \textbf{\bibinfo{volume}{456}},
  \bibinfo{pages}{624} (\bibinfo{year}{2008}).

\bibitem[{\citenamefont{Mannhart et~al.}(2008)\citenamefont{Mannhart, Blank,
  Hwang, Millis, and Triscone}}]{Mannhart08}
\bibinfo{author}{\bibfnamefont{J.}~\bibnamefont{Mannhart}},
  \bibinfo{author}{\bibfnamefont{D.~A.} \bibnamefont{Blank}},
  \bibinfo{author}{\bibfnamefont{H.~Y.} \bibnamefont{Hwang}},
  \bibinfo{author}{\bibfnamefont{A.~J.} \bibnamefont{Millis}},
  \bibnamefont{and} \bibinfo{author}{\bibfnamefont{J.~M.}
  \bibnamefont{Triscone}}, \bibinfo{journal}{MRS Bulletin}
  \textbf{\bibinfo{volume}{33}}, \bibinfo{pages}{1027} (\bibinfo{year}{2008}).

\bibitem[{\citenamefont{Thiel et~al.}(2006)\citenamefont{Thiel, Hammerl,
  Schmehl, Schneider, and Mannhart}}]{Thiel06}
\bibinfo{author}{\bibfnamefont{S.}~\bibnamefont{Thiel}},
  \bibinfo{author}{\bibfnamefont{G.}~\bibnamefont{Hammerl}},
  \bibinfo{author}{\bibfnamefont{A.}~\bibnamefont{Schmehl}},
  \bibinfo{author}{\bibfnamefont{C.~W.} \bibnamefont{Schneider}},
  \bibnamefont{and} \bibinfo{author}{\bibfnamefont{J.}~\bibnamefont{Mannhart}},
  \bibinfo{journal}{Science} \textbf{\bibinfo{volume}{313}},
  \bibinfo{pages}{1942} (\bibinfo{year}{2006}).

\bibitem[{\citenamefont{Yoshimatsu et~al.}(2010)\citenamefont{Yoshimatsu,
  Okabe, Kumigashira, Okamoto, Aizaki, Fujimori, and Oshima}}]{Yoshimatsu10}
\bibinfo{author}{\bibfnamefont{K.}~\bibnamefont{Yoshimatsu}},
  \bibinfo{author}{\bibfnamefont{T.}~\bibnamefont{Okabe}},
  \bibinfo{author}{\bibfnamefont{H.}~\bibnamefont{Kumigashira}},
  \bibinfo{author}{\bibfnamefont{S.}~\bibnamefont{Okamoto}},
  \bibinfo{author}{\bibfnamefont{S.}~\bibnamefont{Aizaki}},
  \bibinfo{author}{\bibfnamefont{A.}~\bibnamefont{Fujimori}}, \bibnamefont{and}
  \bibinfo{author}{\bibfnamefont{M.}~\bibnamefont{Oshima}},
  \bibinfo{journal}{Physical Review Letters} \textbf{\bibinfo{volume}{104}},
  \bibinfo{pages}{147601} (\bibinfo{year}{2010}).

\bibitem[{\citenamefont{Perucchi et~al.}(2010)\citenamefont{Perucchi,
  Baldassarre, Nucara, Calvani, Adamo, Schlom, Orgiani, Maritato, and
  Lupi}}]{Perucchi10}
\bibinfo{author}{\bibfnamefont{A.}~\bibnamefont{Perucchi}},
  \bibinfo{author}{\bibfnamefont{L.}~\bibnamefont{Baldassarre}},
  \bibinfo{author}{\bibfnamefont{A.}~\bibnamefont{Nucara}},
  \bibinfo{author}{\bibfnamefont{P.}~\bibnamefont{Calvani}},
  \bibinfo{author}{\bibfnamefont{C.}~\bibnamefont{Adamo}},
  \bibinfo{author}{\bibfnamefont{D.~G.} \bibnamefont{Schlom}},
  \bibinfo{author}{\bibfnamefont{P.}~\bibnamefont{Orgiani}},
  \bibinfo{author}{\bibfnamefont{L.}~\bibnamefont{Maritato}}, \bibnamefont{and}
  \bibinfo{author}{\bibfnamefont{S.}~\bibnamefont{Lupi}},
  \bibinfo{journal}{Nano Letters} \textbf{\bibinfo{volume}{10}},
  \bibinfo{pages}{4819} (\bibinfo{year}{2010}).

\bibitem[{\citenamefont{Son et~al.}(2010)\citenamefont{Son, Moetakef, LeBeau,
  Ouellette, Balents, Allen, and Stemmer}}]{Son10}
\bibinfo{author}{\bibfnamefont{J.}~\bibnamefont{Son}},
  \bibinfo{author}{\bibfnamefont{P.}~\bibnamefont{Moetakef}},
  \bibinfo{author}{\bibfnamefont{J.~M.} \bibnamefont{LeBeau}},
  \bibinfo{author}{\bibfnamefont{D.}~\bibnamefont{Ouellette}},
  \bibinfo{author}{\bibfnamefont{L.}~\bibnamefont{Balents}},
  \bibinfo{author}{\bibfnamefont{S.~J.} \bibnamefont{Allen}}, \bibnamefont{and}
  \bibinfo{author}{\bibfnamefont{S.}~\bibnamefont{Stemmer}},
  \bibinfo{journal}{Applied Physics Letters} \textbf{\bibinfo{volume}{96}}
  (\bibinfo{year}{2010}).

\bibitem[{\citenamefont{L\"uders et~al.}(2009)\citenamefont{L\"uders, Sheets,
  David, Prellier, and Fr\'esard}}]{Luders09}
\bibinfo{author}{\bibfnamefont{U.}~\bibnamefont{L\"uders}},
  \bibinfo{author}{\bibfnamefont{W.~C.} \bibnamefont{Sheets}},
  \bibinfo{author}{\bibfnamefont{A.}~\bibnamefont{David}},
  \bibinfo{author}{\bibfnamefont{W.}~\bibnamefont{Prellier}}, \bibnamefont{and}
  \bibinfo{author}{\bibfnamefont{R.}~\bibnamefont{Fr\'esard}},
  \bibinfo{journal}{Physical Review B} \textbf{\bibinfo{volume}{80}},
  \bibinfo{pages}{241102} (\bibinfo{year}{2009}).

\bibitem[{\citenamefont{Ariando et~al.}(2011)\citenamefont{Ariando, Wang,
  Baskaran, Liu, Huijben, Yi, Annadi, Barman, Rusydi, Dhar et~al.}}]{Ariando11}
\bibinfo{author}{\bibnamefont{Ariando}},
  \bibinfo{author}{\bibfnamefont{X.}~\bibnamefont{Wang}},
  \bibinfo{author}{\bibfnamefont{G.}~\bibnamefont{Baskaran}},
  \bibinfo{author}{\bibfnamefont{Z.~Q.} \bibnamefont{Liu}},
  \bibinfo{author}{\bibfnamefont{J.}~\bibnamefont{Huijben}},
  \bibinfo{author}{\bibfnamefont{J.~B.} \bibnamefont{Yi}},
  \bibinfo{author}{\bibfnamefont{A.}~\bibnamefont{Annadi}},
  \bibinfo{author}{\bibfnamefont{A.~R.} \bibnamefont{Barman}},
  \bibinfo{author}{\bibfnamefont{A.}~\bibnamefont{Rusydi}},
  \bibinfo{author}{\bibfnamefont{S.}~\bibnamefont{Dhar}}, \bibnamefont{et~al.},
  \bibinfo{journal}{Nat Commun} \textbf{\bibinfo{volume}{2}},
  \bibinfo{pages}{188} (\bibinfo{year}{2011}).

\bibitem[{\citenamefont{Brinkman et~al.}(2007)\citenamefont{Brinkman, Huijben,
  van Zalk, Huijben, Zeitler, Maan, van~der Wiel, Rijnders, Blank, and
  Hilgenkamp}}]{Brinkman07}
\bibinfo{author}{\bibfnamefont{A.}~\bibnamefont{Brinkman}},
  \bibinfo{author}{\bibfnamefont{M.}~\bibnamefont{Huijben}},
  \bibinfo{author}{\bibfnamefont{M.}~\bibnamefont{van Zalk}},
  \bibinfo{author}{\bibfnamefont{J.}~\bibnamefont{Huijben}},
  \bibinfo{author}{\bibfnamefont{U.}~\bibnamefont{Zeitler}},
  \bibinfo{author}{\bibfnamefont{J.~C.} \bibnamefont{Maan}},
  \bibinfo{author}{\bibfnamefont{W.~G.} \bibnamefont{van~der Wiel}},
  \bibinfo{author}{\bibfnamefont{G.}~\bibnamefont{Rijnders}},
  \bibinfo{author}{\bibfnamefont{D.~H.~A.} \bibnamefont{Blank}},
  \bibnamefont{and}
  \bibinfo{author}{\bibfnamefont{H.}~\bibnamefont{Hilgenkamp}},
  \bibinfo{journal}{Nat Mater} \textbf{\bibinfo{volume}{6}},
  \bibinfo{pages}{493} (\bibinfo{year}{2007}).

\bibitem[{\citenamefont{Bert et~al.}(2011)\citenamefont{Bert, Kalisky, Bell,
  Kim, Hikita, Hwang, and Moler}}]{Bert11}
\bibinfo{author}{\bibfnamefont{J.~A.} \bibnamefont{Bert}},
  \bibinfo{author}{\bibfnamefont{B.}~\bibnamefont{Kalisky}},
  \bibinfo{author}{\bibfnamefont{C.}~\bibnamefont{Bell}},
  \bibinfo{author}{\bibfnamefont{M.}~\bibnamefont{Kim}},
  \bibinfo{author}{\bibfnamefont{Y.}~\bibnamefont{Hikita}},
  \bibinfo{author}{\bibfnamefont{H.~Y.} \bibnamefont{Hwang}}, \bibnamefont{and}
  \bibinfo{author}{\bibfnamefont{K.~A.} \bibnamefont{Moler}},
  \bibinfo{journal}{Nat Phys} \textbf{\bibinfo{volume}{7}},
  \bibinfo{pages}{767} (\bibinfo{year}{2011}).

\bibitem[{\citenamefont{Reyren et~al.}(2007)\citenamefont{Reyren, Thiel,
  Caviglia, Kourkoutis, Hammerl, Richter, Schneider, Kopp, Rüetschi, Jaccard
  et~al.}}]{Reyren07}
\bibinfo{author}{\bibfnamefont{N.}~\bibnamefont{Reyren}},
  \bibinfo{author}{\bibfnamefont{S.}~\bibnamefont{Thiel}},
  \bibinfo{author}{\bibfnamefont{A.~D.} \bibnamefont{Caviglia}},
  \bibinfo{author}{\bibfnamefont{L.~F.} \bibnamefont{Kourkoutis}},
  \bibinfo{author}{\bibfnamefont{G.}~\bibnamefont{Hammerl}},
  \bibinfo{author}{\bibfnamefont{C.}~\bibnamefont{Richter}},
  \bibinfo{author}{\bibfnamefont{C.~W.} \bibnamefont{Schneider}},
  \bibinfo{author}{\bibfnamefont{T.}~\bibnamefont{Kopp}},
  \bibinfo{author}{\bibfnamefont{A.-S.} \bibnamefont{Rüetschi}},
  \bibinfo{author}{\bibfnamefont{D.}~\bibnamefont{Jaccard}},
  \bibnamefont{et~al.}, \bibinfo{journal}{Science}
  \textbf{\bibinfo{volume}{317}}, \bibinfo{pages}{1196} (\bibinfo{year}{2007}).

\bibitem[{\citenamefont{Zhang et~al.}(2013)\citenamefont{Zhang, Hwang,
  Raghavan, and Stemmer}}]{Zhang13}
\bibinfo{author}{\bibfnamefont{J.~Y.} \bibnamefont{Zhang}},
  \bibinfo{author}{\bibfnamefont{J.}~\bibnamefont{Hwang}},
  \bibinfo{author}{\bibfnamefont{S.}~\bibnamefont{Raghavan}}, \bibnamefont{and}
  \bibinfo{author}{\bibfnamefont{S.}~\bibnamefont{Stemmer}},
  \bibinfo{journal}{Phys. Rev. Lett.} \textbf{\bibinfo{volume}{110}},
  \bibinfo{pages}{256401} (\bibinfo{year}{2013}).

\bibitem[{\citenamefont{Moetakef
  et~al.}(2012{\natexlab{a}})\citenamefont{Moetakef, Ouellette, Williams,
  James~Allen, Balents, Goldhaber-Gordon, and Stemmer}}]{Moetakef12-2}
\bibinfo{author}{\bibfnamefont{P.}~\bibnamefont{Moetakef}},
  \bibinfo{author}{\bibfnamefont{D.~G.} \bibnamefont{Ouellette}},
  \bibinfo{author}{\bibfnamefont{J.~R.} \bibnamefont{Williams}},
  \bibinfo{author}{\bibfnamefont{S.}~\bibnamefont{James~Allen}},
  \bibinfo{author}{\bibfnamefont{L.}~\bibnamefont{Balents}},
  \bibinfo{author}{\bibfnamefont{D.}~\bibnamefont{Goldhaber-Gordon}},
  \bibnamefont{and} \bibinfo{author}{\bibfnamefont{S.}~\bibnamefont{Stemmer}},
  \bibinfo{journal}{Applied Physics Letters} \textbf{\bibinfo{volume}{101}},
  \bibinfo{eid}{151604} (\bibinfo{year}{2012}{\natexlab{a}}).

\bibitem[{\citenamefont{Chen et~al.}(2013)\citenamefont{Chen, Lee, and
  Balents}}]{Chen13}
\bibinfo{author}{\bibfnamefont{R.}~\bibnamefont{Chen}},
  \bibinfo{author}{\bibfnamefont{S.}~\bibnamefont{Lee}}, \bibnamefont{and}
  \bibinfo{author}{\bibfnamefont{L.}~\bibnamefont{Balents}},
  \bibinfo{journal}{Phys. Rev. B} \textbf{\bibinfo{volume}{87}},
  \bibinfo{pages}{161119} (\bibinfo{year}{2013}).

\bibitem[{\citenamefont{Bjaalie et~al.}(2014)\citenamefont{Bjaalie, Janotti,
  Himmetoglu, and Van~de Walle}}]{Bjaalie14}
\bibinfo{author}{\bibfnamefont{L.}~\bibnamefont{Bjaalie}},
  \bibinfo{author}{\bibfnamefont{A.}~\bibnamefont{Janotti}},
  \bibinfo{author}{\bibfnamefont{B.}~\bibnamefont{Himmetoglu}},
  \bibnamefont{and} \bibinfo{author}{\bibfnamefont{C.~G.} \bibnamefont{Van~de
  Walle}}, \bibinfo{journal}{Phys. Rev. B} \textbf{\bibinfo{volume}{90}},
  \bibinfo{pages}{195117} (\bibinfo{year}{2014}).

\bibitem[{\citenamefont{Gunnarsson et~al.}(2003)\citenamefont{Gunnarsson,
  Calandra, and Han}}]{Gunnarsson03}
\bibinfo{author}{\bibfnamefont{O.}~\bibnamefont{Gunnarsson}},
  \bibinfo{author}{\bibfnamefont{M.}~\bibnamefont{Calandra}}, \bibnamefont{and}
  \bibinfo{author}{\bibfnamefont{J.~E.} \bibnamefont{Han}},
  \bibinfo{journal}{Rev. Mod. Phys.} \textbf{\bibinfo{volume}{75}},
  \bibinfo{pages}{1085} (\bibinfo{year}{2003}).

\bibitem[{\citenamefont{Hussey et~al.}(2004)\citenamefont{Hussey, Takenaka, and
  Takagi}}]{Hussey04}
\bibinfo{author}{\bibfnamefont{N.~E.} \bibnamefont{Hussey}},
  \bibinfo{author}{\bibfnamefont{K.}~\bibnamefont{Takenaka}}, \bibnamefont{and}
  \bibinfo{author}{\bibfnamefont{H.}~\bibnamefont{Takagi}},
  \bibinfo{journal}{Philosophical Magazine} \textbf{\bibinfo{volume}{84}},
  \bibinfo{pages}{2847} (\bibinfo{year}{2004}).

\bibitem[{\citenamefont{Kresse and Furthm\"uller}(1996)}]{Kresse96}
\bibinfo{author}{\bibfnamefont{G.}~\bibnamefont{Kresse}} \bibnamefont{and}
  \bibinfo{author}{\bibfnamefont{J.}~\bibnamefont{Furthm\"uller}},
  \bibinfo{journal}{Phys. Rev. B} \textbf{\bibinfo{volume}{54}},
  \bibinfo{pages}{11169} (\bibinfo{year}{1996}).

\bibitem[{\citenamefont{Kresse and Joubert}(1999)}]{Kresse99}
\bibinfo{author}{\bibfnamefont{G.}~\bibnamefont{Kresse}} \bibnamefont{and}
  \bibinfo{author}{\bibfnamefont{D.}~\bibnamefont{Joubert}},
  \bibinfo{journal}{Phys. Rev. B} \textbf{\bibinfo{volume}{59}},
  \bibinfo{pages}{1758} (\bibinfo{year}{1999}).

\bibitem[{\citenamefont{Liechtenstein et~al.}(1995)\citenamefont{Liechtenstein,
  Anisimov, and Zaanen}}]{Liechtenstein95}
\bibinfo{author}{\bibfnamefont{A.~I.} \bibnamefont{Liechtenstein}},
  \bibinfo{author}{\bibfnamefont{V.~I.} \bibnamefont{Anisimov}},
  \bibnamefont{and} \bibinfo{author}{\bibfnamefont{J.}~\bibnamefont{Zaanen}},
  \bibinfo{journal}{Phys. Rev. B} \textbf{\bibinfo{volume}{52}},
  \bibinfo{pages}{R5467} (\bibinfo{year}{1995}).

\bibitem[{\citenamefont{Unoki and Sakudo}(1967)}]{Unoki67}
\bibinfo{author}{\bibfnamefont{H.}~\bibnamefont{Unoki}} \bibnamefont{and}
  \bibinfo{author}{\bibfnamefont{T.}~\bibnamefont{Sakudo}},
  \bibinfo{journal}{Journal of the Physical Society of Japan}
  \textbf{\bibinfo{volume}{23}}, \bibinfo{pages}{546} (\bibinfo{year}{1967}).

\bibitem[{\citenamefont{Okamoto et~al.}(2006)\citenamefont{Okamoto, Millis, and
  Spaldin}}]{Okamoto06}
\bibinfo{author}{\bibfnamefont{S.}~\bibnamefont{Okamoto}},
  \bibinfo{author}{\bibfnamefont{A.~J.} \bibnamefont{Millis}},
  \bibnamefont{and} \bibinfo{author}{\bibfnamefont{N.~A.}
  \bibnamefont{Spaldin}}, \bibinfo{journal}{Phys. Rev. Lett.}
  \textbf{\bibinfo{volume}{97}}, \bibinfo{pages}{056802}
  (\bibinfo{year}{2006}).

\bibitem[{\citenamefont{Pentcheva and Pickett}(2006)}]{Pentcheva06}
\bibinfo{author}{\bibfnamefont{R.}~\bibnamefont{Pentcheva}} \bibnamefont{and}
  \bibinfo{author}{\bibfnamefont{W.~E.} \bibnamefont{Pickett}},
  \bibinfo{journal}{Phys. Rev. B} \textbf{\bibinfo{volume}{74}},
  \bibinfo{pages}{035112} (\bibinfo{year}{2006}).

\bibitem[{\citenamefont{Pentcheva and Pickett}(2007)}]{Pentcheva07}
\bibinfo{author}{\bibfnamefont{R.}~\bibnamefont{Pentcheva}} \bibnamefont{and}
  \bibinfo{author}{\bibfnamefont{W.~E.} \bibnamefont{Pickett}},
  \bibinfo{journal}{Phys. Rev. Lett.} \textbf{\bibinfo{volume}{99}},
  \bibinfo{pages}{016802} (\bibinfo{year}{2007}).

\bibitem[{\citenamefont{Son et~al.}(2009)\citenamefont{Son, Cho, Lee, Lee, and
  Han}}]{Son209}
\bibinfo{author}{\bibfnamefont{W.-j.} \bibnamefont{Son}},
  \bibinfo{author}{\bibfnamefont{E.}~\bibnamefont{Cho}},
  \bibinfo{author}{\bibfnamefont{B.}~\bibnamefont{Lee}},
  \bibinfo{author}{\bibfnamefont{J.}~\bibnamefont{Lee}}, \bibnamefont{and}
  \bibinfo{author}{\bibfnamefont{S.}~\bibnamefont{Han}},
  \bibinfo{journal}{Physical Review B} \textbf{\bibinfo{volume}{79}},
  \bibinfo{pages}{245411} (\bibinfo{year}{2009}).

\bibitem[{\citenamefont{Delugas et~al.}(2011)\citenamefont{Delugas, Filippetti,
  Fiorentini, Bilc, Fontaine, and Ghosez}}]{Delugas11}
\bibinfo{author}{\bibfnamefont{P.}~\bibnamefont{Delugas}},
  \bibinfo{author}{\bibfnamefont{A.}~\bibnamefont{Filippetti}},
  \bibinfo{author}{\bibfnamefont{V.}~\bibnamefont{Fiorentini}},
  \bibinfo{author}{\bibfnamefont{D.~I.} \bibnamefont{Bilc}},
  \bibinfo{author}{\bibfnamefont{D.}~\bibnamefont{Fontaine}}, \bibnamefont{and}
  \bibinfo{author}{\bibfnamefont{P.}~\bibnamefont{Ghosez}},
  \bibinfo{journal}{Phys. Rev. Lett.} \textbf{\bibinfo{volume}{106}},
  \bibinfo{pages}{166807} (\bibinfo{year}{2011}).

\bibitem[{\citenamefont{Moetakef
  et~al.}(2012{\natexlab{b}})\citenamefont{Moetakef, Jackson, Hwang, Balents,
  Allen, and Stemmer}}]{Moetakef12}
\bibinfo{author}{\bibfnamefont{P.}~\bibnamefont{Moetakef}},
  \bibinfo{author}{\bibfnamefont{C.~A.} \bibnamefont{Jackson}},
  \bibinfo{author}{\bibfnamefont{J.}~\bibnamefont{Hwang}},
  \bibinfo{author}{\bibfnamefont{L.}~\bibnamefont{Balents}},
  \bibinfo{author}{\bibfnamefont{S.~J.} \bibnamefont{Allen}}, \bibnamefont{and}
  \bibinfo{author}{\bibfnamefont{S.}~\bibnamefont{Stemmer}},
  \bibinfo{journal}{Phys. Rev. B} \textbf{\bibinfo{volume}{86}},
  \bibinfo{pages}{201102} (\bibinfo{year}{2012}{\natexlab{b}}).

\bibitem[{\citenamefont{Benedek et~al.}(2012)\citenamefont{Benedek, Mulder, and
  Fennie}}]{Benedek12}
\bibinfo{author}{\bibfnamefont{N.~A.} \bibnamefont{Benedek}},
  \bibinfo{author}{\bibfnamefont{A.~T.} \bibnamefont{Mulder}},
  \bibnamefont{and} \bibinfo{author}{\bibfnamefont{C.~J.}
  \bibnamefont{Fennie}}, \bibinfo{journal}{Journal of Solid State Chemistry}
  \textbf{\bibinfo{volume}{195}}, \bibinfo{pages}{11 } (\bibinfo{year}{2012}).

\bibitem[{\citenamefont{Meyer et~al.}(1999)\citenamefont{Meyer, Padilla, and
  Vanderbilt}}]{Meyer99}
\bibinfo{author}{\bibfnamefont{B.}~\bibnamefont{Meyer}},
  \bibinfo{author}{\bibfnamefont{J.}~\bibnamefont{Padilla}}, \bibnamefont{and}
  \bibinfo{author}{\bibfnamefont{D.}~\bibnamefont{Vanderbilt}},
  \bibinfo{journal}{Faraday Discuss.} \textbf{\bibinfo{volume}{114}},
  \bibinfo{pages}{395} (\bibinfo{year}{1999}).

\bibitem[{\citenamefont{Nakagawa et~al.}(2006)\citenamefont{Nakagawa, Hwang,
  and Muller}}]{Nakagawa06}
\bibinfo{author}{\bibfnamefont{N.}~\bibnamefont{Nakagawa}},
  \bibinfo{author}{\bibfnamefont{H.~Y.} \bibnamefont{Hwang}}, \bibnamefont{and}
  \bibinfo{author}{\bibfnamefont{D.~A.} \bibnamefont{Muller}},
  \bibinfo{journal}{Nat Mater} \textbf{\bibinfo{volume}{5}},
  \bibinfo{pages}{204} (\bibinfo{year}{2006}).

\bibitem[{\citenamefont{Millis and Schlom}(2010)}]{Millis10}
\bibinfo{author}{\bibfnamefont{A.~J.} \bibnamefont{Millis}} \bibnamefont{and}
  \bibinfo{author}{\bibfnamefont{D.~G.} \bibnamefont{Schlom}},
  \bibinfo{journal}{Phys. Rev. B} \textbf{\bibinfo{volume}{82}},
  \bibinfo{pages}{073101} (\bibinfo{year}{2010}).

\bibitem[{\citenamefont{Chen et~al.}(2010)\citenamefont{Chen, Kolpak, and
  Ismail-Beigi}}]{Chen10}
\bibinfo{author}{\bibfnamefont{H.}~\bibnamefont{Chen}},
  \bibinfo{author}{\bibfnamefont{A.}~\bibnamefont{Kolpak}}, \bibnamefont{and}
  \bibinfo{author}{\bibfnamefont{S.}~\bibnamefont{Ismail-Beigi}},
  \bibinfo{journal}{Physical Review B} \textbf{\bibinfo{volume}{82}},
  \bibinfo{pages}{085430} (\bibinfo{year}{2010}).

\bibitem[{\citenamefont{Park and Millis}(2013)}]{SP}
\bibinfo{author}{\bibfnamefont{S.~Y.} \bibnamefont{Park}} \bibnamefont{and}
  \bibinfo{author}{\bibfnamefont{A.~J.} \bibnamefont{Millis}},
  \bibinfo{journal}{Phys. Rev. B} \textbf{\bibinfo{volume}{87}},
  \bibinfo{pages}{205145} (\bibinfo{year}{2013}).

\bibitem[{\citenamefont{Lee et~al.}(2013)\citenamefont{Lee, Podraza, Zhu,
  Berger, Shen, Sestak, Collins, Kourkoutis, Mundy, Wang et~al.}}]{Lee13}
\bibinfo{author}{\bibfnamefont{C.-H.} \bibnamefont{Lee}},
  \bibinfo{author}{\bibfnamefont{N.~J.} \bibnamefont{Podraza}},
  \bibinfo{author}{\bibfnamefont{Y.}~\bibnamefont{Zhu}},
  \bibinfo{author}{\bibfnamefont{R.~F.} \bibnamefont{Berger}},
  \bibinfo{author}{\bibfnamefont{S.}~\bibnamefont{Shen}},
  \bibinfo{author}{\bibfnamefont{M.}~\bibnamefont{Sestak}},
  \bibinfo{author}{\bibfnamefont{R.~W.} \bibnamefont{Collins}},
  \bibinfo{author}{\bibfnamefont{L.~F.} \bibnamefont{Kourkoutis}},
  \bibinfo{author}{\bibfnamefont{J.~A.} \bibnamefont{Mundy}},
  \bibinfo{author}{\bibfnamefont{H.}~\bibnamefont{Wang}}, \bibnamefont{et~al.},
  \bibinfo{journal}{Applied Physics Letters} \textbf{\bibinfo{volume}{102}},
  \bibinfo{eid}{122901} (\bibinfo{year}{2013}).

\end{thebibliography}
\end{document}